\documentclass[fleqn,usenatbib]{mnras}

\DeclareRobustCommand{\VAN}[3]{#2}
\let\VANthebibliography\thebibliography
\def\thebibliography{\DeclareRobustCommand{\VAN}[3]{##3}\VANthebibliography}

\usepackage{graphicx} 
\usepackage{amsmath}
\usepackage{mathtools}
\usepackage{newtxtext,newtxmath}
\usepackage{tabularx}
\usepackage{array}
\usepackage{url}
\usepackage{makecell}
\usepackage{tikz}
\usepackage{orcidlink}

\newcommand{\halomass}{$M_{\mathrm{200c}}$}
\newcommand{\virialradius}{$R_{\mathrm{200c}}$}
\newcommand{\stellarmass}{$M_{\star}$}
\newcommand{\solarmass}{M$_{\sun}$}
\title[Physical and observable properties of the CGM]{The halo mass dependence of physical and observable properties in the circumgalactic medium}

\author[A. W. S. Cook et al.]{
Andrew W. S. Cook\textsuperscript{\orcidlink{https://orcid.org/0009-0001-9328-6903}},$^{1}$\thanks{E-mail: cookaw@cardiff.ac.uk}
Freeke van de Voort\textsuperscript{\orcidlink{https://orcid.org/0000-0002-6301-638X}},$^{1}$
Rüdiger Pakmor\textsuperscript{\orcidlink{https://orcid.org/0000-0003-3308-2420}}$^{2}$
 and Robert J. J. Grand\textsuperscript{\orcidlink{https://orcid.org/0000-0001-9667-1340}}$^{3}$
\\
$^{1}$Cardiff Hub for Astrophysics Research and Technology, School of Physics and Astronomy, Cardiff University, Queens Building, The Parade, Cardiff, CF24 3AA, UK\\
$^{2}$Max-Planck-Institut für Astrophysik, Karl-Schwarzschild-Str. 1, D-85748, Garching, Germany\\
$^{3}$Astrophysics Research Institute, Liverpool John Moores University, 146 Brownlow Hill, Liverpool, L3 5RF, UK
}

\date{Accepted XXX. Received YYY; in original form ZZZ}

\pubyear{\the\year{}}

\begin{document}
\label{firstpage}
\pagerange{\pageref{firstpage}--\pageref{lastpage}}
\maketitle


\begin{abstract}
We study the dependence of the physical and observable properties of the circumgalactic medium on its halo mass. We analyse 22 simulations from the Auriga suite of high resolution cosmological magneto-hydrodynamical `zoom-in' simulations. We focus on the current epoch ($z=0$) and halo masses between $10^{10}~\text{\solarmass}\leq\text{\halomass}\leq10^{12}$~\solarmass. The median temperature and metallicity increase with halo mass as expected. We find a larger scatter in temperature and smaller scatter in metallicity at higher halo masses. The scatter of temperature and metallicity as a function of radius is increases at larger radii. The median and scatter of the volume-weighted density and the mass-weighted radial velocity shows no significant dependence on halo mass. Our results highlight that the CGM is more multiphase in haloes at higher halo masses. We additionally investigate column densities for \ion{H}{I} and the metal ions \ion{C}{IV}, \ion{O}{VI}, \ion{Mg}{II} and \ion{Si}{II} as a function of stellar mass and galactocentric radius. We find that the \ion{H}{I} and metal ion column densities increase with the stellar mass of the system at sufficiently large radii ($R\gtrsim{0.2}$\virialradius). We find good agreement between our \ion{H}{I} column densities and observations outside $20$\% of the virial radius and overpredict \ion{H}{I} within $20$\%. \ion{Mg}{II} and \ion{Si}{II} are similarly overpredicted within $20$\% of the virial radius, but drop off steeply at larger radii. Our \ion{O}{VI} column densities underpredict observations for stellar masses between $10^{9.7}~\text{\solarmass}\leq\text{\stellarmass}<10^{10.8}$~\solarmass\space with good agreement at $10^{10.8}$~\solarmass. \ion{C}{IV} column densities agree with observational detections above a halo mass of $10^{9.7}$\solarmass. We find that \ion{O}{VI} (\ion{Mg}{II}) traces the highest (lowest) temperatures, and lowest (highest) density and metallicity. OVI and CIV are photo-ionized (collisionally ionized) at low (high) halo masses with a transition to higher temperatures at $10^{11}$~\solarmass. However, there is no clear trend for the radial velocity of the ions. These results demonstrate similarities and discrepancies between our simulations of Milky Way mass haloes and observations, as well as highlighting that further constraints are needed in less massive haloes.

\end{abstract}



\section{Introduction}

The circumgalactic medium (CGM) is the gaseous component of dark matter-dominated haloes. It surrounds galaxies and contains gas accreted from the intergalactic medium (IGM) and gas ejected from the interstellar medium (ISM). The CGM plays a major role in the evolution of galaxies, acting as a reservoir for baryonic matter that can accrete on to the ISM and provide fuel for star formation \citep{Tumlinson2017}.\par
The CGM features various temperature phases which range from $10^{4}$~K (or lower if we account for molecular outflows) up to approximately the virial temperature (or potentially hotter in outflows), with different phases in approximate pressure equilibrium. These gas phases are probed observationally primarily through absorption line spectroscopy as it is sensitive to the low densities in the CGM, making it an effective method of characterising the low-density CGM. Absorption line spectroscopy traces ions with different ionization potentials to detect various gas temperatures, and thus the multiphase nature, of the CGM \citep{Lehner2013,Anand2021,Mathur2021}.\par
Quantifying the physical nature of the CGM is important to further our understanding of its role in galaxy evolution. We know from simulations and analytical arguments that the hydrodynamical properties of the CGM depend not only on feedback from the galaxy, but also on the halo mass \citep[e.g.][]{Frenk1988,Wang2008,vandevoort2012}. Haloes with masses similar to the Milky Way ($10^{12}$~\solarmass) are typically hot ($\mathrm{T}\approx10^{6}$~K) and metal-rich with higher accretion rates on to the galaxy than lower mass haloes \citep{Keres2005,Tumlinson2011}.\par
Feedback from galaxies, driven by stellar winds, supernovae and active galactic nuclei (AGN), eject matter into the CGM which enriches the environment with metals \citep{Muratov2015,Johnson2017,Sanchez2019}. The strong feedback from star-forming galaxies means that only $20-25\%$ of all metals produced are retained in their ISM \citep{Peeples2014}. Material ejected by these feedback processes either flows out of the CGM into the IGM and beyond, remains in the CGM, or re-accretes on to the galaxy. This re-accretion is likely important at late times as it can dominate the accretion rate \citep{Oppenheimer2010,Hafen2020,Wright2021} and could be important for maintaining star formation in the galaxy.\par
Outflows from feedback also change the ionization states of metals by heating the gas. Low ionization states such as \ion{H}{I}, \ion{Si}{II} and \ion{Mg}{II} trace cool gas ($\mathrm{T}=10^{4}$~K) in the CGM, with \ion{C}{IV} tracing warmer gas of $10^{4.5-5}$K, and higher ionization states such as \ion{O}{VI} tracing warm-hot temperatures, if in collisional ionization equilibrium, of $10^{5.5}$~K \citep{Gnat2007,Wiersma2009,Strawn2023}.\par
The ionization states of gas in the CGM can be probed in simulations too, utilising radiative transfer or non-equilibrium chemistry models \citep[e.g. \textsc{arepo-rt} and \textsc{ramses-rtz:}][]{Kannan2019,Katz2022}{}{} or post-processing tools such as \textsc{trident} \citep{Hummels2017} which uses \textsc{cloudy} \citep{Ferland2017} for its ionization tables. Mock observables have been compared to observations before with varying results. \citet{Machado2018} show over predictions in the \ion{O}{VI} and \ion{Si}{III} column densities when compared to observations. Studies also show good agreement for synthetic absorption lines and equivalent widths, and covering fractions from \textsc{simba}  consistent with observations \citep{Appleby2021}. The aim of these comparisons is to assist in constraining our models, thus leading to better predictions in future simulation work.\par
\begin{figure}
\centering
\includegraphics[width=0.5\textwidth]{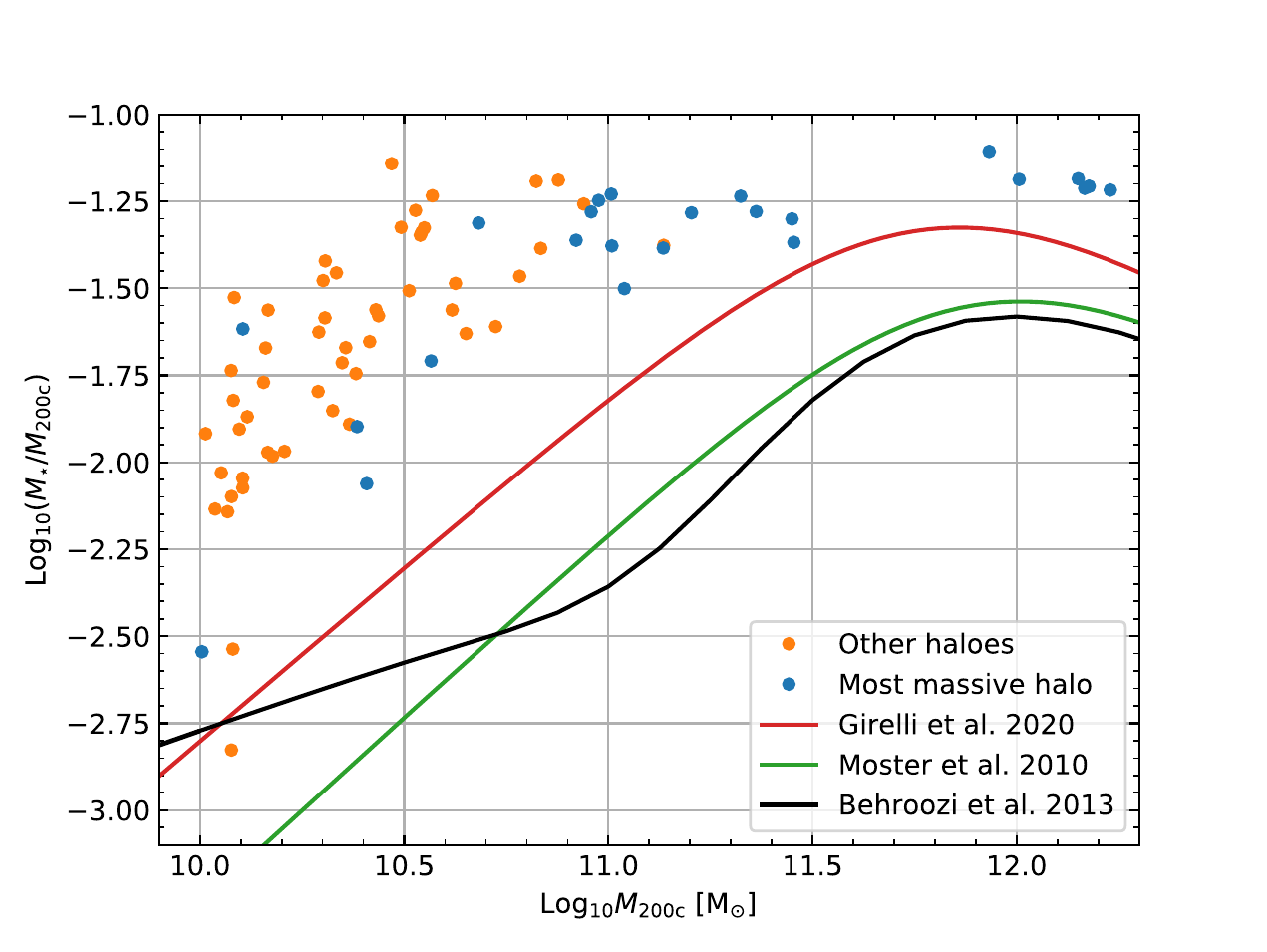}
\vspace{-15pt}
\caption{Stellar mass to halo mass (SM-HM) ratio against halo mass. Blue data points show the SM-HM ratio for the most massive halo from each simulation. We compare them to three models from \protect\citealt{Moster2010}, \protect\citealt{Behroozi2013}, and \protect\citealt{Girelli2020} shown as green, black, and red curves respectively. Compared to these models, our simulations over-predict the SM-HM ratio but generally follow the same trend.}
\label{fig:stel_halo}
\end{figure}
Observationally, little is known about the physical nature of the CGM around dwarf galaxies. While certain observations have been successful in measuring column densities and some properties of the CGM around dwarf galaxies \citep[e.g.][]{Bordoloi2014,Burchett2015,Johnson2017,Zheng2020,Tchernyshyov2022,Zheng2023} the underlying physical nature is difficult to constrain. It is therefore important that we constrain the physical nature of the CGM of dwarf galaxies using cosmological simulations \citep{Peeples2019}.\par
In this work, we analyse cosmological zoom-in simulations to determine the dependence of the temperature, density, metallicity, and radial velocity on halo mass for haloes with a mass range of $10^{10}$~\solarmass$\leq$\halomass$\leq10^{12}$~\solarmass. The aim of this work is to better understand the CGM in a cosmological context by quantifying the scatter in the temperature, density, metallicity and radial velocity as a function of halo mass and galactocentric radius. Furthermore, we compare the column densities for \ion{H}{I}, \ion{C}{IV}, \ion{O}{VI}, \ion{Mg}{II} and \ion{Si}{II} in our simulations with observational surveys including COS-Halos \citep{Tumlinson2013, Werk2016}, COS-Dwarfs \citep{Bordoloi2014} as well as other observational studies \citep{Johnson2015, Zheng2023} to better understand how our model compares to the existing observations and find discrepancies that we can use to inform future models.\par
This paper is structured as follows. We briefly outline the Auriga simulations, including how these haloes are selected in Sec.~2. We will then discuss the physical properties of the CGM in our simulated haloes in Sec.~3. We compare the results of our column densities to those from observations in Sec.~4. Finally, we conclude and discuss our results in Sec.~5.

\section{The Auriga Suite of Cosmological Simulations}

\begin{figure*}
\centering
\includegraphics[width=0.96\textwidth]{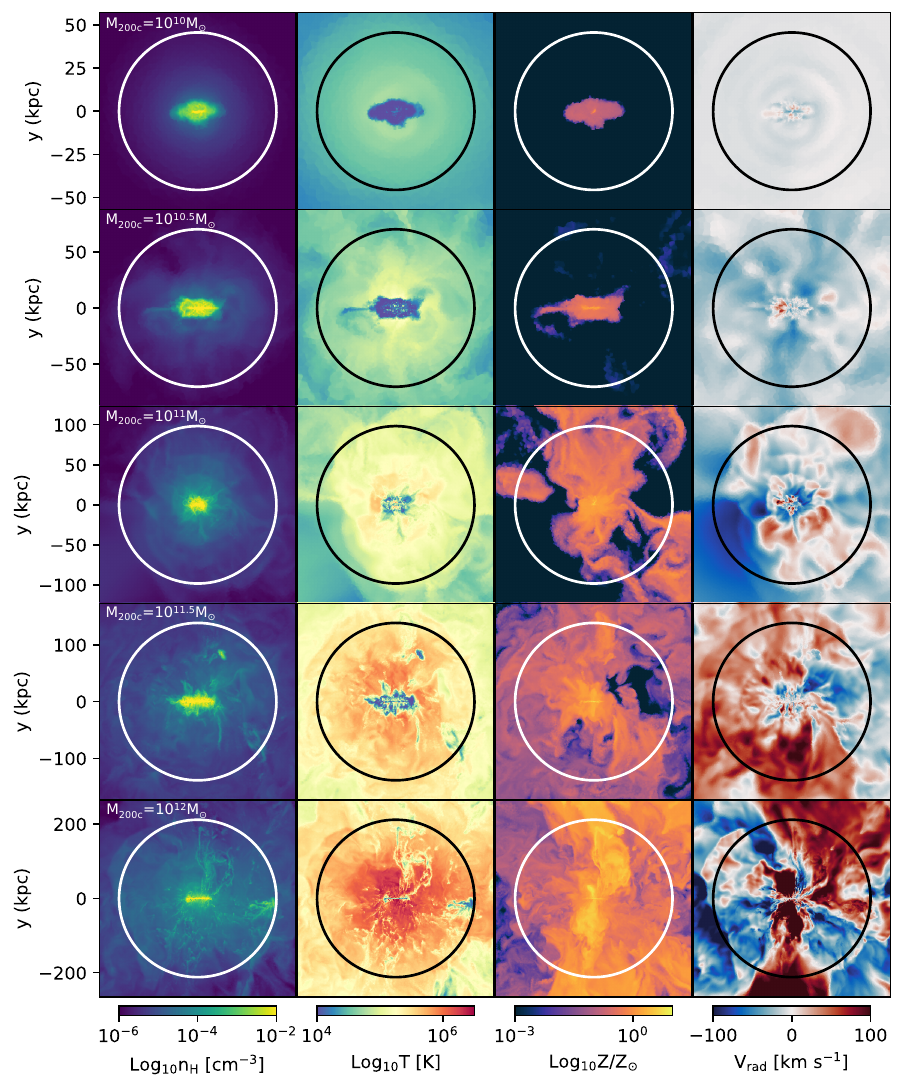}
\vspace{-12pt}
\caption{$2.5$\virialradius$\times2.5$\virialradius\space thin projections of five haloes, rotated such that the stellar disc is edge-on. Columns show, from left to right, the volume-weighted hydrogen number density and mass-weighted temperature, metallicity, and radial velocity with a projection depth of \virialradius$/8$. The halo mass increases from top to bottom from $10^{10}$~\solarmass$-10^{12}$~\solarmass\space in steps of $0.5$ dex. The circle in each panel indicates the virial radius of each halo. When compared to high-mass haloes, low-mass haloes (\halomass$<10^{11}$~\solarmass) show smaller scatter in the temperature, the density, the metallicity and the radial velocity. The metallicity of the low-mass haloes is highest in the centre but is still several orders of magnitude lower than the solar values of high-mass haloes with a sharp and steep drop off in metallicity outside of the centre. The radial velocities are also significantly lower.}
\label{fig:collated_images}
\end{figure*}
This work studies the CGM of 22 high-resolution simulations from the Auriga project -- a suite of magneto-hydrodynamical (MHD) cosmological zoom-in simulations \citep{Grand2017,Grand2024}. The simulations we use have a baryonic mass resolution of ${\sim}6\times10^{3}$~\solarmass\space (also referred to as 'level 3') and a dark matter resolution of ${\sim}4\times10^{4}$~\solarmass.\par
We focus on gas within the virial radius (\virialradius) of the halo, defined as the radius within which the density within it is approximately $200$ times the critical density of the universe. We study haloes within a halo mass range of $10^{10}$~\solarmass$\leq$\halomass$\leq10^{12}$~\solarmass, corresponding to a stellar mass range of $10^{7.5}$~\solarmass$\leq$\stellarmass$\leq10^{11}$~\solarmass. For six simulations, the most massive halo at the centre of the zoom region has a halo mass between $10^{10}$~\solarmass$-10^{11}$~\solarmass, 12 between $10^{11}$~\solarmass$-10^{11.5}$~\solarmass\space and six between $1\times10^{12}$~\solarmass$-2\times10^{12}$~\solarmass. Table \ref{table:1} in the Appendix lists further information on the key properties of the most massive haloes in each of the 22 zoom-in simulations. The following is a brief overview of the simulations and a description of the post-processing method to obtain metal ion column densities.\par
Figure \ref{fig:stel_halo} shows the ratio of stellar mass to halo mass as a function of halo mass. This includes the 22 most massive haloes and 64 additional haloes within the zoom-in region of each simulation. We find our simulations overestimate the stellar mass based on abundance matching from \citet{Moster2010}, \citet{Girelli2020} and \citet{Behroozi2013}. The larger scatter of lower mass haloes has also been found in other simulation suites with an explicit ISM model \citep{Agertz20,Gutcke2021,Sales22}.\par
The initial conditions of Auriga are based on the dark matter-only simulations from the EAGLE project \citep{Schaye2015}, with a co-moving box size of $100$ Mpc. Haloes are identified using a friends-of-friends (FoF) algorithm with a standard linking length \citep{Davis1985}. Haloes are selected based on their halo mass then by how isolated those haloes are, of which the halo selected for zoom is randomly selected from the lowest percentile. The mass selection criterion for Milky Way mass haloes is $0.5<$\halomass$/10^{12}$~\solarmass$<2$ and less massive haloes were selected within $0.5<$\halomass$/10^{11}$~\solarmass$<5$ and $0.5<$\halomass$/10^{10}$~\solarmass$<5$. The high resolution zoom-in region extends to $\approx5$\virialradius\space of each halo. Auriga adopts cosmological parameters $\Omega_{m}=0.307$, $\Omega_{b}=0.048$ and $\Omega_{\Lambda}=0.693$ with $H_{0}= 100h\,\mathrm{km\,s}^{-1}$ Mpc$^{-1}$ where $h=0.6777$ from \citet{Planck2014}. \par
The simulations were run from $z=127$ to $z=0$ using the arbitrary Lagrangian-Eulerian moving-mesh code \textsc{arepo} \citep{Springel2010}, which includes magneto-hydrodynamics (MHD) for gas and collisionless dynamics (for dark matter, stars and black holes). \textsc{arepo} uses an unstructured Voronoi mesh with idealised MHD solved by a finite volume second order Runge-Kutta integration scheme \citep{Pakmor2016a}.\par
The Auriga galaxy formation model is described fully in \citet{Grand2017} and summarised here. The ISM subgrid model in these simulations is described in \citet{Springel2003} as a two-phase medium of dense, cold gas clouds surrounded by hotter, ambient gas. The gas in the ISM is assumed to be star-forming once the density exceeds a threshold density of $n_{\mathrm{H}}^{*}=0.13~\mathrm{cm^{-3}}$. These simulations additionally include primordial and metal line-cooling \citep{vogel2013} as well as an ultraviolet background with self-shielding \citep{Giguere2009,Rahmati2013}.\par
Auriga incorporates feedback from supernovae, stellar winds and AGN. Stellar feedback is implemented through gas cells that are temporarily converted into collisionless wind particles, ejected from star-forming gas, until they reach 5\% of $n_{\mathrm{H}}^{*}$ where they recouple and deposit mass, momentum, thermal energy and metals. The minimum temperature in our simulations is set to $10^{4}$K because cooling below this temperature is inefficient.\par
\begin{figure*}
\centering
\includegraphics[width=
\textwidth]{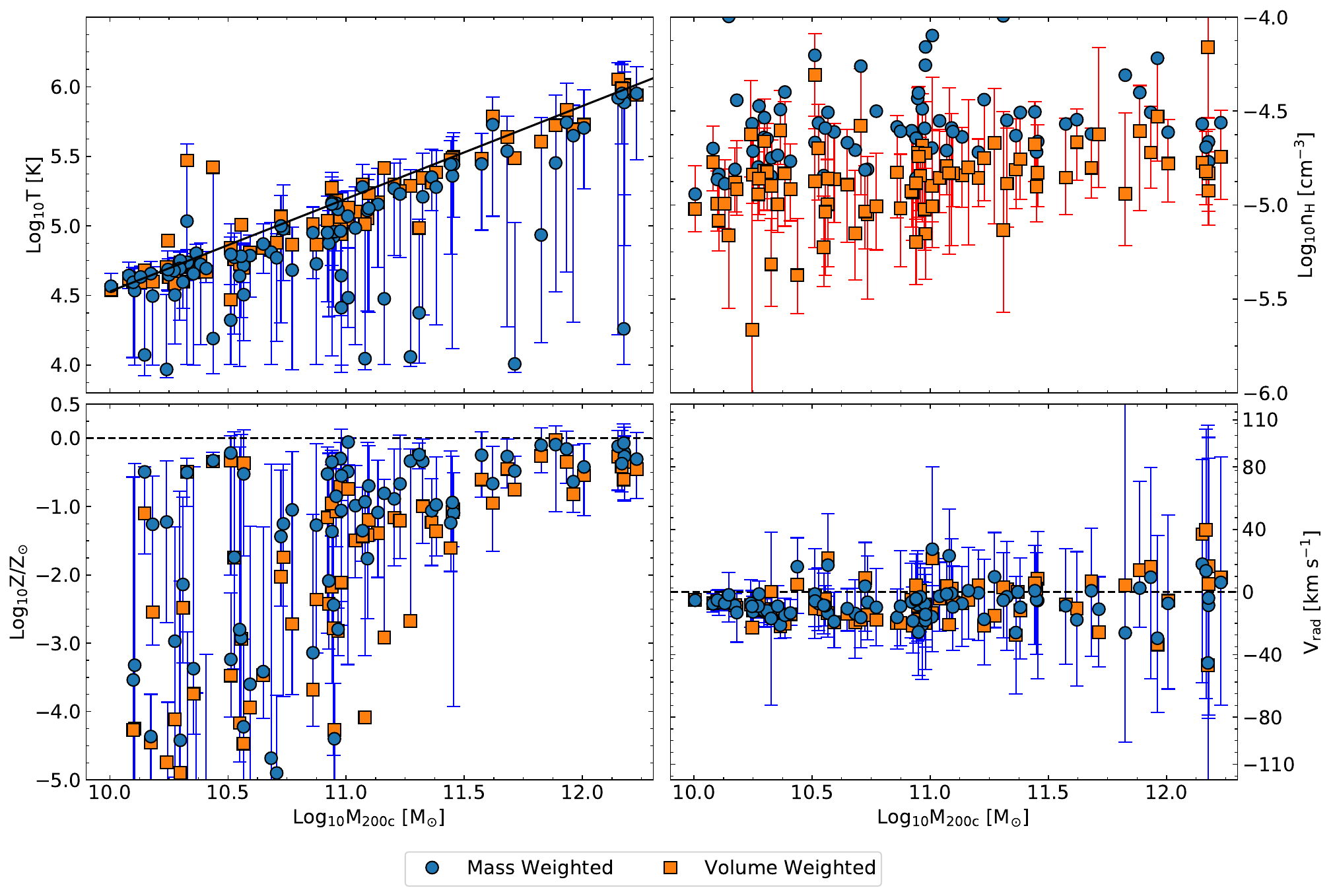}
\vspace{-12pt}
\caption{Median CGM temperature (top, left), hydrogen number density (top, right), metallicity (bottom, left) and radial velocity (bottom, right) for 86 haloes within the zoom-in region of our simulations as a function of halo mass with scatter bars showing the 16$^{\mathrm{th}}$ and 84$^{\mathrm{th}}$ percentiles. The median shown is volume-weighted (orange squares) for density and mass-weighted (blue circles) for the other properties. The scatter is only shown for mass-weighted temperature, metallicity and radial velocity and for the volume-weighted density. We additionally include the virial temperature from \protect\citet{Mo2010} in the temperature panel. The properties are measured between $0.3\leq{R}$/\virialradius$\leq1.0$ to remove the central galaxy and its extended \ion{H}{I}-dominated disc. The temperature and metallicity vary strongly for our haloes, both in their median values and their scatter. There is no clear dependence on halo mass for the density. The scatter in radial velocity is wider in more massive haloes. However, the median radial velocity shows no clear trend with halo mass and is in most cases dominated by inflows.}
\label{fig:med_properties}
\end{figure*}
AGN feedback is modelled as a two-mode process: one which heats the gas locally around the black hole and the other which gently heats bubbles of gas at randomly placed locations up to $80$\% of the virial radius. Black holes are seeded in the simulations with a mass of $10^{5}$~$h^{-1}$~\solarmass\space when the FoF halo group has a mass $M_\mathrm{FOF}=5\times10^{10}$~$h^{-1}$~\solarmass. Ionising radiation is also included in the AGN feedback model in the form of X-rays.\par
We calculate the column density of \ion{H}{I}, \ion{C}{IV}, \ion{O}{VI}, \ion{Mg}{II} and \ion{Si}{II} in our simulations. We chose these ions because they are the most commonly observed ions through absorption line spectroscopy \citep{Fox2005,Ranjan_2022} and trace different temperature phases of the gas. The mass fraction of \ion{H}{I} is calculated on-the-fly whereas the mass fraction of the metal ions are calculated in post-processing. We use tables generated with \textsc{cloudy} \citep{Ferland2017} by \citet{Hummels2017} to compute the mass fraction of the ions. We can then use this ion fraction and the metal specific mass fraction to calculate the column density along any sight-line in our simulations.\par
A resolution test was conducted comparing this set of 22 simulations with different mass resolutions: ${\sim}8\times10^{2}$~\solarmass\space (also referred to as 'level 2') and ${\sim}5.4\times10^{4}$~\solarmass\space (also referred to as 'level 4'). The level 2 simulations all had halo masses $\leq10^{10.5}$~\solarmass\space and the level 4 simulations had halo masses $\geq10^{11}$~\solarmass.
\begin{figure*}
\centering
\includegraphics[width=\textwidth]{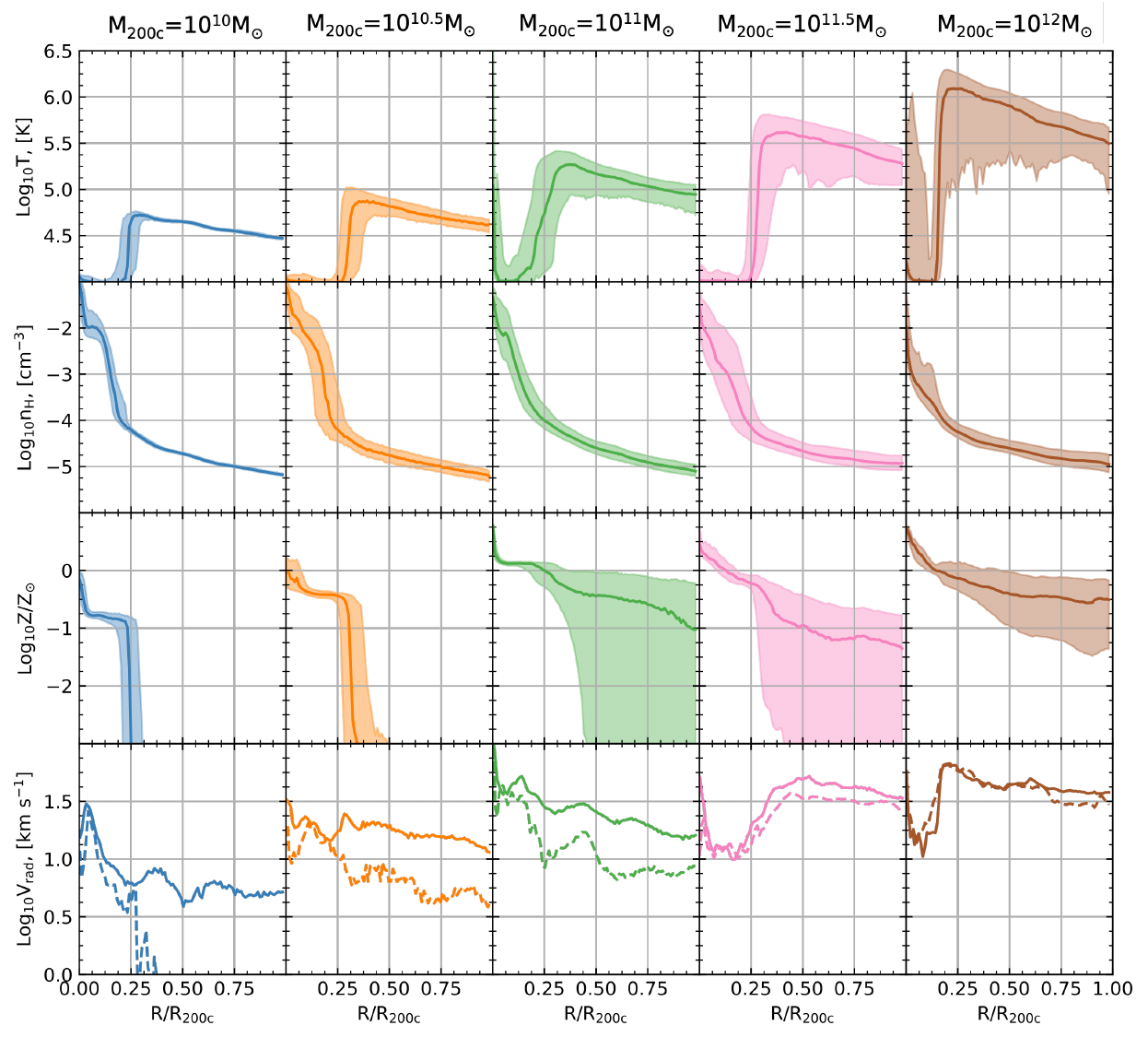}
\vspace{-12pt}
\caption{Median physical properties from top to bottom: temperature, density, metallicity and radial velocity as a function of radius normalised by the virial radius. Shaded regions show the $16^{\mathrm{th}}$ and $84^{\mathrm{th}}$ percentile of the properties as a measure of the scatter for the haloes from Fig.\ref{fig:collated_images}. The halo mass increases from left to right, from $10^{10}$~\solarmass$-10^{12}$~\solarmass\space in steps of $0.5$ dex. The temperature, metallicity and radial velocity are mass-weighted whereas the density is volume-weighted. Gas associated with satellites and the ISM has been excluded. The temperature profile decreases at $R>0.3$\virialradius\space with a steeper decrease in more massive haloes. The metallicity is highest in the centre in all haloes. Haloes with \halomass$\leq10^{10.5}$~\solarmass\space have a steep drop off at radii larger than $0.25$\virialradius. For \halomass$>10^{10.5}$~\solarmass, the metallicity profile is much shallower and the scatter is large. Reasonably metal-rich gas can be found in the outer CGM as well as metal-poor gas. Similar to Figure \ref{fig:med_properties}, the density shows no strong dependence on halo mass, with a steep drop off in the inner $0.25$\virialradius\space CGM and a more shallow profile in the outer CGM. Finally, the outflows in \halomass$\leq10^{11}$~\solarmass\space haloes is significantly weaker than the inflowing matter. In more massive haloes, the inflows and outflows are at approximately equal rates.}
\label{fig:radial_properties}
\end{figure*}

\section{Dependence of Physical Properties in the CGM on Halo Mass}
One of the most difficult components of the CGM to observationally constrain is the physical properties including its temperature, density, metallicity and radial velocity. While the temperature can be inferred, the CGM properties are largely unconstrained, especially in low-mass haloes. We use our cosmological simulations to quantify the temperature, density, metallicity and radial velocity as a function of halo mass and radius to obtain a broad understanding of the physical nature of the CGM in our simulations.
\subsection{Physical Nature of the CGM}
Figure \ref{fig:collated_images} shows thin projections at $z=0$ of the volume-weighted density and mass-weighted temperature, metallicity, and radial velocity for five haloes rotated such that the central galaxy is edge-on. The halo mass ranges from $10^{10}$~\solarmass$-10^{12}$~\solarmass, increasing in mass intervals of $0.5$ dex from top to bottom. The circles indicate the virial radius of each halo: $45$~kpc, $70$~kpc, $98$~kpc, $139$~kpc, and $212$~kpc, from top to bottom.\par
For more massive haloes, the CGM is hotter, more enriched with metals, and has higher radial velocities, both inflowing (negative values) and outflowing (positive values). In our most massive halo (bottom row) bipolar outflows can be seen along the minor axis which have high metallicity and radial velocities but show no significant temperature differences along this axis. High values along the minor axis are seen in radial velocity and metallicity but are lower along the major axis. The temperature and density typically decrease with radius with minor differences outside of the central $0.25$\virialradius.\par
In contrast, our lowest mass haloes (top row) exhibit low scatter in their temperature, density, metallicity and radial velocity. High values of each property are found in the centre of the halo. Because the radial velocity is low and the gas is mostly inflowing, metals cannot reach significant distances into the CGM, resulting in a deficit in metals in the outer CGM. Similarly the temperature of the CGM does not change significantly outside of the central $0.25$\virialradius. Temperatures reach ${\sim}10^{4.5}$~K at most and do not fluctuate much at fixed radii.\par
Figure \ref{fig:med_properties} shows the density, temperature, metallicity and radial velocity as a function of halo mass for 86 haloes within the zoom-in region of our 22 simulations. We calculate the median and $16^{\mathrm{th}}$ and $84^{\mathrm{th}}$ percentile scatter for the gas within $0.3$\virialradius$-1.0$\virialradius\space to exclude the dense, cool gas in the centre. The blue circles show mass-weighted values and the orange squares show volume-weighted values. For clarity, the scatter is only shown for mass-weighted temperature, metallicity and radial velocity and for the volume-weighted density. We additionally include the virial temperature from \citet{Mo2010} in the temperature panel.\par
As expected, we find that the median temperature is higher in more massive haloes, increasing from ${\sim}10^{4.0}$~K in our lowest mass haloes (\halomass$=10^{10}$~\solarmass) up to temperatures of ${\sim}10^{6}$~K at high masses (\halomass$=10^{12}$~\solarmass). Furthermore, the scatter is wider at these high masses, with a range of ${\sim}0.5$ dex in low-mass haloes, increasing to ${\sim}1.5$ dex in high-mass haloes. This increase indicates that the CGM is more multiphase in more massive haloes. The $16^{\mathrm{th}}$ percentile of most of our haloes above \halomass$>10^{11.75}$~\solarmass\space haloes are at temperatures above $10^{4}$~K. Cool gas is still present just not in large quantities.\par
The gas density shows no clear dependence on the halo mass. The scatter fluctuates by $<0.1$ dex from halo to halo, but shows no clear trend with halo mass for either mass- or volume-weighted quantities. The scatter seen here is primarily caused by the decreasing trend in the average density from $0.3$\virialradius\space to \virialradius\space (see Figure~\ref{fig:radial_properties}).\par
The median metallicity has no clear dependence on halo mass for \halomass $<10^{11}$~\solarmass\space (low-mass haloes), with more metal-rich gas present in more massive haloes; median values cover a range from ${\sim}10^{-5}-10^{-1}$ in low-mass haloes to near solar values (${\sim}10^{0}$) in more massive haloes (\halomass $\geq10^{11}$~\solarmass). The scatter is small in more massive haloes, decreasing from ${\sim}4$ ($\leq10^{-5}-10^{-0.5}$) dex to ${\sim}1$ dex ($10^{-1}-10^{0}$) over our sample of haloes. We find that the most massive haloes in the zoom-in region, which were selected to be the most isolated, exhibit low metallicities. It is therefore possible that additional haloes located within the zoom-in region have gained metals from nearby systems. The stronger feedback from more massive galaxies produces stronger outflows, therefore contributing to more metal enrichment in the outer CGM of more massive haloes.\par
The median radial velocity is about $-40-40$~km/s across the halo mass range. Inflows dominate slightly over outflows, but shows no clear dependence on halo mass. The scatter in radial velocity is more skewed towards inflows in less massive haloes (\halomass$<10^{11.5}$~\solarmass) while more massive haloes show more balanced inflows and outflows.\par
The relationship between volume- and mass-weighted quantities changes for different properties. The volume- and mass-weighted values of temperature and radial velocity are similar. The volume-weighted metallicity is typically lower than the mass-weighted metallicity in low-mass haloes (\halomass$<10^{11}$~\solarmass) though the difference is smaller in more massive haloes indicating metals are clumped in higher density gas.
\begin{figure*}
\centering
\includegraphics[width=\textwidth]{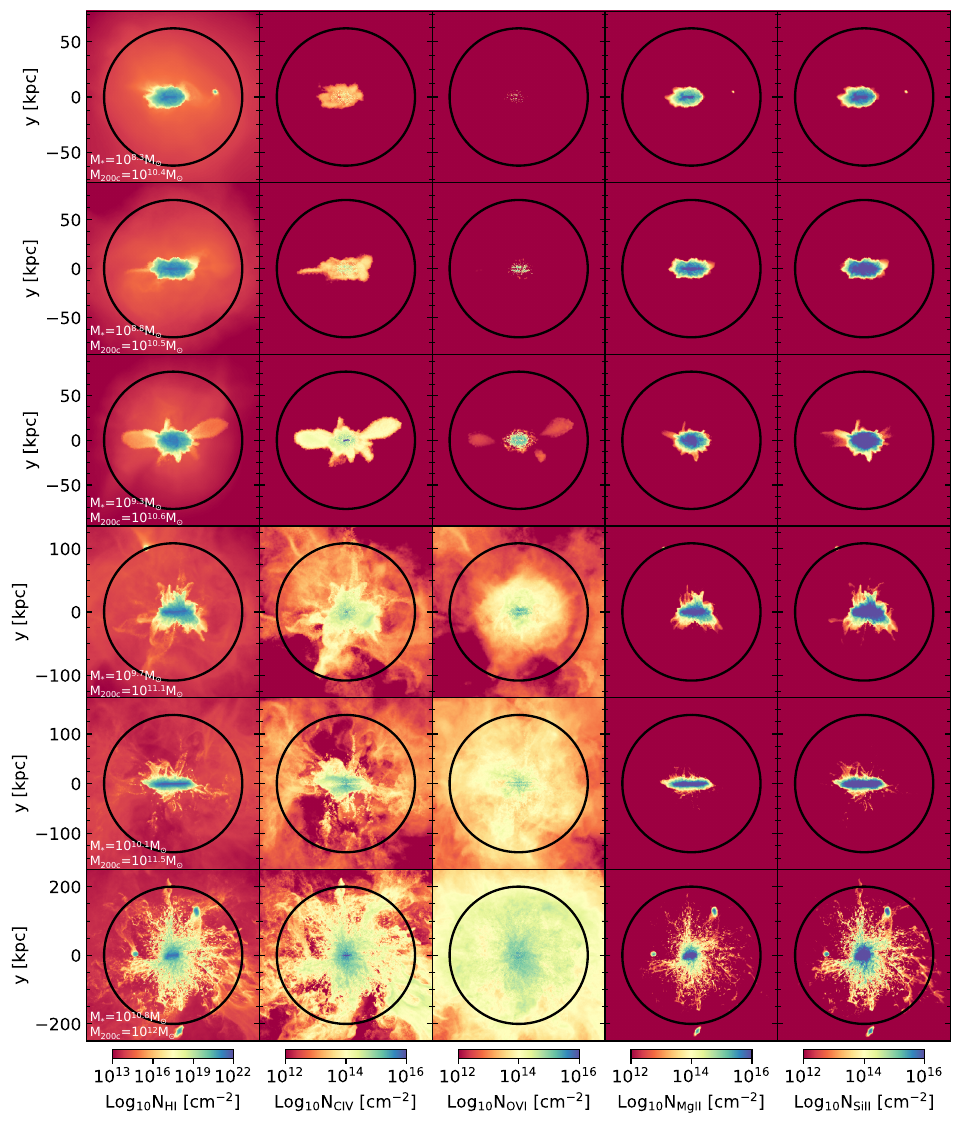}
\vspace{-12pt}
\caption{$2.5$\virialradius$\times2.5$\virialradius\space projections of six haloes with a depth of $2$\virialradius, rotated such that the stellar disc is edge-on. The chosen haloes are selected based on the stellar mass of the central galaxies and are thus different from those in Figure \ref{fig:collated_images}. From top to bottom the stellar masses increase from $10^{8.3}$~\solarmass, $0^{8.8}$~\solarmass, $10^{9.3}$~\solarmass, $10^{9.7}$~\solarmass, $10^{10.1}$~\solarmass\space to $10^{10.8}$~\solarmass. From left to right, columns correspond to \ion{H}{I}, \ion{C}{IV}, \ion{O}{VI}, \ion{Mg}{II} and \ion{Si}{II}. The virial radius is indicated by the black circle in each panel. Column densities are highest in the centres of all haloes and decrease with radius. $N_{\mathrm{\ion{C}{IV}}}$ and $N_{\mathrm{\ion{O}{VI}}}$ increase with stellar mass, reaching high values even in the outer CGM.}
\label{fig:col_dens_image}
\end{figure*}
\subsection{Radial Profiles}
Figure \ref{fig:radial_properties} shows the radial dependence of the temperature, density, metallicity and radial velocity, for both inflowing (solid line) and outflowing (dashed line) gas, for the same five haloes as shown in Figure \ref{fig:collated_images}. The solid curves show the median properties and the shaded region the $16^{\mathrm{th}}$ and $84^{\mathrm{th}}$ percentile scatter. The haloes shown range in halo mass from $10^{10}$~\solarmass$-10^{12}$~\solarmass, increasing from left to right in steps of 0.5 dex. Properties are mass-weighted except density which is volume-weighted. We exclude star-forming gas from this figure.\par
The median temperature remains fairly constant at $10^{4}$~K in the inner CGM (typically between $0.1-0.25$\virialradius\space with some variation between haloes). This region is dominated by a non-star-forming extended gas disc. As matter accretes onto the CGM, the temperature has been shown to increase to values near the virial temperature \citep{vandevoort2012}. When the gas reaches higher densities in the inner CGM, it is able to cool and join the central disc. The scatter in temperature of the gas in the outer CGM is larger in more massive haloes, indicating that the gas in the outer CGM (R$>0.25$\virialradius) is more multiphase compared to less massive haloes.\par
The highest densities ($10^{-1}\mathrm{cm}^{-3}$) are found in the centre of the CGM, typically within $0.25$\virialradius, decreasing down to $10^{-5}\mathrm{cm}^{-3}$ at the virial radius. The scatter is most notable in our $10^{10}$~\solarmass\space haloes with very low scatter outside of $0.25$\virialradius. The scatter above this halo mass is generally consistent for all other haloes. We see a radial dependence in the median profile but no change in scatter or halo mass dependence on gas density when volume-weighted. We separately calculated the mass-weighted density which showed the same median trend as volume-weighted density. The density remains high at ${\sim}10^{-1}\mathrm{cm}^{-2}$ out to larger radii than volume-weighted, typically out to $0.25$\virialradius, before a sharp drop in density. Additionally, mass-weighted density shows somewhat higher scatter in more massive haloes.\par
The highest metallicity is seen in the centre of our haloes in the inner ${\sim}0.25$\virialradius. For our low-mass dwarfs, \halomass\space$<10^{11}$~\solarmass, the metallicity decreases sharply because the outflows do not reach larger distances. The central metallicity is $1$ dex lower in the $10^{10}$~\solarmass\space halo compared to the $10^{11}$~\solarmass\space halo. For more massive haloes (\halomass\space$\geq 10^{11}$~\solarmass) the median metallicity following a decreasing profile, with the scatter increasing with radius.\par
We see higher inflow and outflow velocities for more massive haloes. Both inflows and outflows follow similar trends within $0.25$\virialradius\space for all haloes. Haloes with \halomass$<10^{11.5}$~\solarmass\space have lower radial velocities in outflowing gas compared to inflowing gas outside of $0.25$\virialradius, and are thus dominated by inflowing gas out to the virial radius. More massive haloes feature faster inflows than low mass haloes and similarly high outflows. Haloes with \halomass$\geq10^{11.5}$~\solarmass\space exhibit inflowing and outflowing gas increasing to a maximum radial velocity at roughly the same radius as the maximum temperature. At this point, the gas is in rotation leading to less inflowing and outflowing velocities.\par

\section{Column Density of the CGM}
\begin{figure*}
\centering
\includegraphics[width=\textwidth]{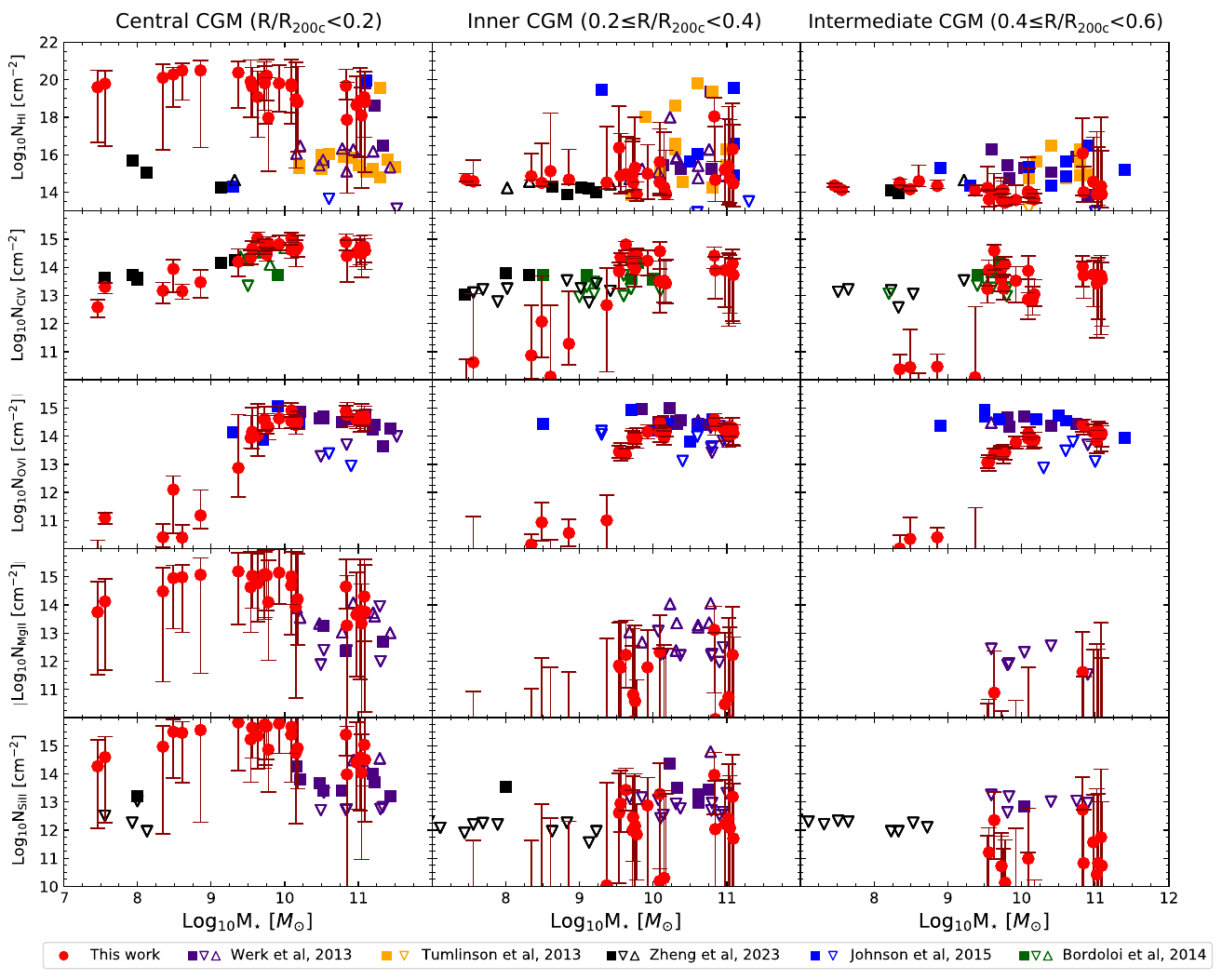}
\vspace{-12pt}
\caption{Simulated and observed column densities as a function of \stellarmass. From top to bottom, red circles show the median of $N_{\mathrm{\ion{H}{I}}}$, $N_{\mathrm{\ion{C}{IV}}}$, $N_{\mathrm{\ion{O}{VI}}}$, $N_{\mathrm{\ion{Mg}{II}}}$ and $N_{\mathrm{\ion{Si}{II}}}$ with $16^{\mathrm{th}}$ and $84^{\mathrm{th}}$ percentile scatter bars. We separate the column densities within $0.6$\virialradius\space in three radial bins of $0.2$\virialradius\space width. The depth of the projection is $2$\virialradius\space and the median is measured along two orthogonal edge-on directions. We compare with observational data from \protect\citet{Werk2013}, \protect\citet{Tumlinson2013}, \protect\citet{Bordoloi2014}, \protect\citet{Johnson2015} and \protect\citet{Zheng2023}. Detections are shown as filled squares and upper (lower) limits as empty downwards (upwards) arrows. The observational error bars are omitted for readability. Additionally, the ISM is removed from our measurements to select only CGM gas of the main halo. The column density of warm ions (\ion{C}{IV} and \ion{O}{VI}) is high above $10^{9.5}$~\solarmass\space stellar mass haloes for all radial bins. \ion{C}{IV} is relatively constant above \stellarmass$=10^{9.5}$~\solarmass\space and agrees with detections at all radii for higher mass haloes and in lower mass haloes within $0.2$\virialradius\space. Lower mass systems have significantly lower column densities, especially at larger radii. \ion{O}{VI} broadly agrees with observations for \stellarmass$>10^{9.5}$~\solarmass\space at multiple radii while lower mass systems show $N_{\mathrm{\ion{O}{VI}}}$ values roughly ${\sim}3$ dex lower. The cool regime (\ion{H}{I}, \ion{Mg}{II} and \ion{Si}{II}) has high column density in the centre for all haloes but drops off significantly above $0.2$\virialradius. $N_{\mathrm{\ion{Mg}{II}}}$ and $N_{\mathrm{\ion{Si}{II}}}$ are higher in systems with \stellarmass$>10^{9.5}$ than in lower mass systems at large radii. \ion{H}{I} has larger scatter in more massive haloes but no significant dependence of the median $N_{\mathrm{\ion{H}{I}}}$ on stellar mass.}
\label{fig:median_col_dens}
\end{figure*}
It is important to understand how our simulations match or differ from observations. This will assist in understanding the validity of our models and how best we can improve them in future simulations.\par
We compare the column density ($N_{x}$ where $x$ is \ion{H}{I}, \ion{C}{IV}, \ion{O}{VI}, \ion{Si}{II} or \ion{Mg}{II}) of \ion{H}{I}, \ion{C}{IV}, \ion{O}{VI}, \ion{Si}{II} and \ion{Mg}{II} in our simulations with observational data from \citet{Bordoloi2014}, \citet{Werk2013}, \citet{Tumlinson2013}, \citet{Johnson2015}, and \citet{Zheng2023}. We exclude observational errors as they are $\approx\pm{0.01-0.15}$ and are therefore negligible compared to the scale of the scatter of our simulations. The CGM is ionized via photo-ionization and collisional ionization. The level of ionization depends on the density and temperature of the gas, as well as the redshift \citep[e.g.][]{Strawn2023}. We define two temperature regimes: the warm regime which is traced by \ion{C}{IV} and \ion{O}{VI} at temperatures of $\mathrm{T}=10^{4.5}$~K$-10^{5.5}$~K, and a cool regime, traced by \ion{H}{I}, \ion{Mg}{II} and \ion{Si}{II} at temperatures of $\mathrm{T}= 10^{4.0}$~K$-10^{4.5}$~K.\par
\begin{figure*}
\centering
\includegraphics[width=\textwidth]{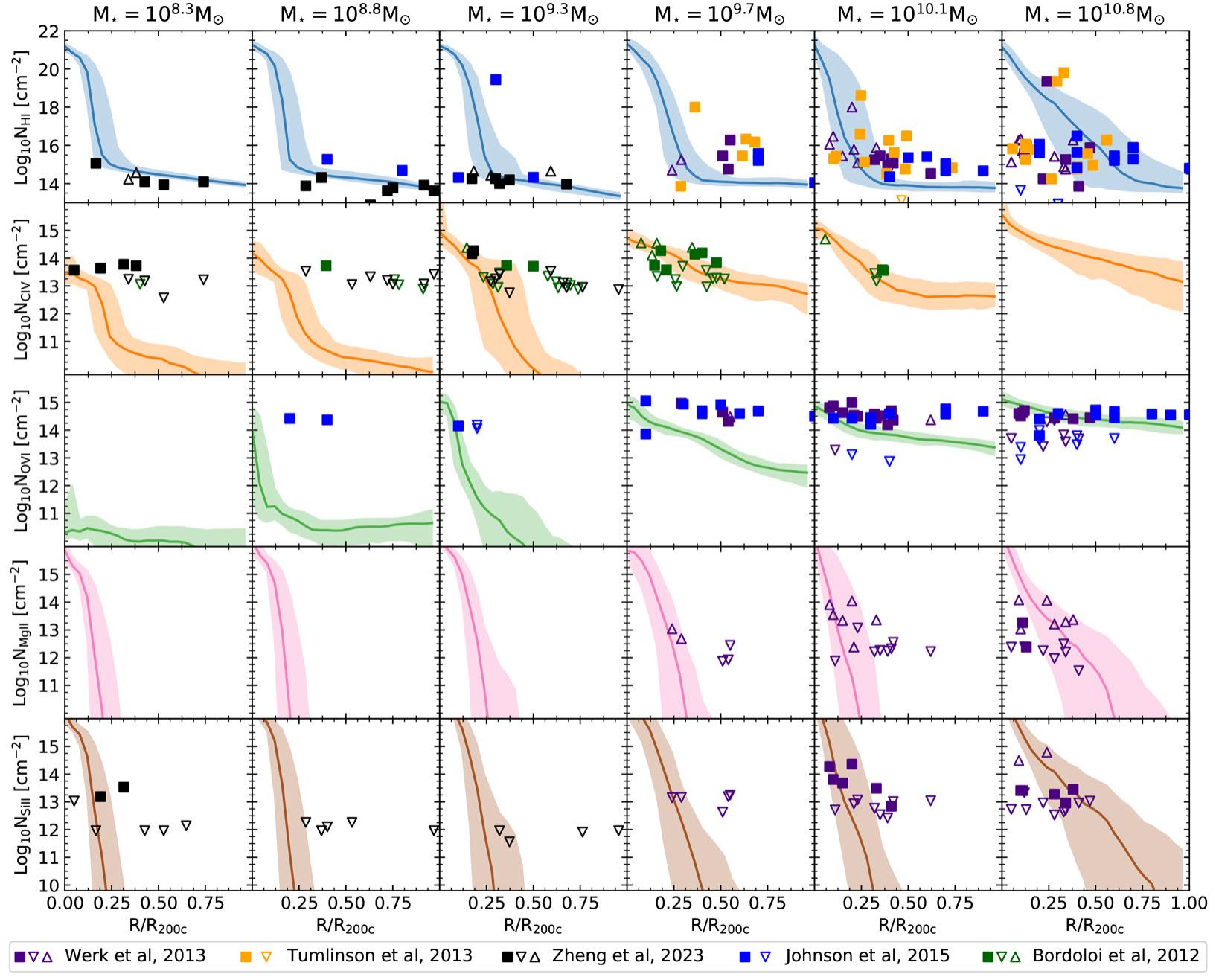}
\vspace{-12pt}
\caption{Radial profiles of \ion{H}{I} and ion column densities, where the radius is normalised by \virialradius, for the same haloes as in Fig.~\ref{fig:col_dens_image}. The median is shown by solid curves and the $16^{\mathrm{th}}$ and $84^{\mathrm{th}}$ percentiles are shown as shaded regions. From top to bottom, each row shows the radial profile for $N_{\mathrm{\ion{H}{I}}}$, $N_{\mathrm{\ion{C}{IV}}}$, $N_{\mathrm{\ion{O}{VI}}}$, $N_{\mathrm{\ion{Mg}{II}}}$ and $N_{\mathrm{\ion{Si}{II}}}$. Our radial profiles are compared with observational data, binned within $10^{8.0}$~\solarmass$<$\stellarmass$<10^{11}$~\solarmass\space in equal intervals of $0.5$ dex, increasing from left to right. Haloes from our simulations are selected to have stellar masses in the middle of these mass bins. Similarly to Figure \ref{fig:median_col_dens}, the ISM is removed from the simulations. The column density of all ions increases with stellar mass in the outer CGM. Our column density for \ion{H}{I} agrees with most observational detections in haloes with \stellarmass$<10^{9.7}$~\solarmass. We see underpredictions at radii $R>0.5$\virialradius\space for $10^{9.7}$~\solarmass\space and $10^{10.1}$~\solarmass, and an overprediction at low radii at \stellarmass$=10^{10.8}$~\solarmass. \ion{C}{IV} agrees with observations at small radii ($R<0.4$\virialradius) for \stellarmass$\geq10^{9.3}$~\solarmass\space while \ion{O}{VI} is slightly under-produced at all stellar masses except the most massive (\stellarmass$=10^{10.8}$~\solarmass). \ion{Mg}{II} and \ion{Si}{II} follow similar radial profiles with sharp drop off between ${\sim}0.2$\virialradius$-0.3$\virialradius. There are few observational detections below $10^{10.1}$~\solarmass\space to compare with our column densities. We do however find reasonable agreement with observations in the two most massive haloes studied.}
\label{fig:col_dens_radial}
\end{figure*}
Figure \ref{fig:col_dens_image} shows halo projections oriented edge-on to the galactic stellar disc of, from left to right, $N_{\mathrm{\ion{H}{I}}}$, $N_{\mathrm{\ion{C}{IV}}}$, $N_{\mathrm{\ion{O}{VI}}}$, $N_{\mathrm{\ion{Mg}{II}}}$ and $N_{\mathrm{\ion{Si}{II}}}$. Six different haloes are chosen based on their stellar mass. The six haloes have stellar masses between $10^{8.3}$~\solarmass$-10^{10.8}$~\solarmass\space increasing from top to bottom in steps of $0.5$ dex. High column densities of \ion{H}{I} and our metal ions are seen in the centres of these haloes. The low ionization state ions highlight the filamentary structure of the outer CGM in the most massive haloes (\stellarmass\space$=10^{10.8}$~\solarmass). We find low $N_{\mathrm{\ion{C}{IV}}}$ and $N_{\mathrm{\ion{O}{VI}}}$ in the outer CGM of galaxies with \stellarmass$\leq10^{9.5}$~\solarmass\space (\halomass\space$\leq10^{11.0}$~\solarmass\space). $N_{\mathrm{\ion{C}{IV}}}$ and $N_{\mathrm{\ion{O}{VI}}}$ are higher in the outer CGM of more massive haloes, because these haloes contain higher temperature gas.\par
\subsection{Column Density and Stellar Mass}
Figure \ref{fig:median_col_dens} shows the various column densities as a function of stellar mass. The median is shown as red circles and the scatter is quantified with $16^{\mathrm{th}}$ and $84^{\mathrm{th}}$ percentiles and shown as scatter bars. The median and percentiles are calculated along two orthogonal axes for which the central galaxy has been rotated to be edge-on. We split these data into three radial bins of $0.0$\virialradius$-0.2$\virialradius\space (central CGM), $0.2$\virialradius$-0.4$\virialradius\space (inner CGM) and $0.4$\virialradius$-0.6$\virialradius\space (intermediate CGM). Observational detections are shown as filled squares, upper limits as open downward arrows, and lower limits as open upward arrows. We additionally set the ion fraction to zero in the ISM.\par
The median and scatter of $N_{\mathrm{\ion{H}{I}}}$ decreases with radius for all stellar masses. In the central CGM, our median $N_{\mathrm{\ion{H}{I}}}$ over-predicts observations though there is significant scatter towards lower column densities. The median $N_{\mathrm{\ion{H}{I}}}$ in the inner CGM decreases to ${\sim}10^{14}\mathrm{cm^{-2}}-{\sim}10^{15}\mathrm{cm^{-2}}$ where we find our simulations agree reasonably well with observations. In the intermediate CGM, most observational data at \stellarmass$>10^{9}$~\solarmass\space are somewhat higher than our median column densities for \ion{H}{I}. However, our haloes also show significant scatter towards higher column densities. Observations also show substantial scatter throughout the CGM, with most $N_{\mathrm{\ion{H}{I}}}$ detections falling within the upper percentile of our scatter above a stellar mass of $10^{9}$~\solarmass.\par
The median $N_{\mathrm{\ion{Mg}{II}}}$ and $N_{\mathrm{\ion{Si}{II}}}$ follow a similar trend to \ion{H}{I} for all stellar masses: an over-prediction compared to the observational data in the centre and a decrease with radius which leads to values more in agreement with observations. Column densities range from $10^{14}\mathrm{cm^{-2}}$ to $10^{16}\mathrm{cm^{-2}}$ in the centre and follow a decreasing radial trend. The median values of $N_{\mathrm{\ion{Si}{II}}}$ fall within the range of observed values and upper limits for \stellarmass$>10^{9.5}$~\solarmass\space in the inner CGM and underpredict observations in the intermediate CGM. The median $N_{\mathrm{\ion{Mg}{II}}}$ is typically ${\sim}1$ dex lower than a systems \ion{Si}{II} and underpredicts observations in both the inner and intermediate CGM. This is partially due to the underproduction of Mg following the yield set of \citep{Portinari1998}. For \stellarmass$\leq10^{9.5}$~\solarmass, $N_{\mathrm{\ion{Mg}{II}}}$ and $N_{\mathrm{\ion{Si}{II}}}$ decrease to values that would be undetectable in observations.\par
We find similar median $N_{\mathrm{\ion{C}{IV}}}$ values to those observed within the central CGM for all stellar masses. $N_{\mathrm{\ion{C}{IV}}}$ increases with stellar mass up to $10^{9.5}$~\solarmass\space beyond which the column density remains constant. The median $N_{\mathrm{\ion{C}{IV}}}$ decreases and its scatter increases with increasing radii. The most significant decrease in median $N_{\mathrm{\ion{C}{IV}}}$ is seen in haloes with \stellarmass$\leq10^{9.5}$~\solarmass. A larger increase in the scatter of $N_{\mathrm{\ion{C}{IV}}}$ occurs in haloes below a stellar mass of $10^{9.5}$~\solarmass. From the central CGM to the inner CGM, our median $N_{\mathrm{\ion{C}{IV}}}$ decreases by ${\sim}2$ dex from $10^{13}~\mathrm{cm}^{-2}-10^{14}~\mathrm{cm}^{-2}$ to $10^{11}~\mathrm{cm}^{-2}-10^{13}~\mathrm{cm}^{-2}$. This is below observational detections for \stellarmass$\leq10^{9.5}$~\solarmass. This decrease also happens in the intermediate CGM by about the same ${\sim}2$ dex, leading to a total decrease of ${\sim}4$ dex. For \stellarmass$>10^{9.5}$~\solarmass, $N_{\mathrm{\ion{C}{IV}}}$ decreases by ${\sim}2$ dex from from the central to the intermediate CGM and largely agrees with observations.\par
Our $N_{\mathrm{\ion{O}{VI}}}$ agrees with observational detections for \stellarmass$>10^{9.5}$~\solarmass\space in the central CGM. The median $N_{\mathrm{\ion{O}{VI}}}$ decreases by ${\sim}1$ dex from the central CGM ($N_{\ion{O}{VI}}=10^{14}-10^{15}~\mathrm{cm}^{-2}$) to the inner CGM ($N_{\ion{O}{VI}}=10^{13}-10^{14}~\mathrm{cm}^{-2}$) at these stellar masses. At larger radii, the systems with \stellarmass$>10^{10.7}$~\solarmass\space have a less significant decrease than systems below $10^{10.7}$~\solarmass\space and remain in agreement with observational detections. For \stellarmass$\leq10^{9.5}$~\solarmass, the median $N_{\mathrm{\ion{O}{VI}}}$ is ${\sim}3-4$ dex lower than in more massive systems. The scatter in $N_{\mathrm{\ion{O}{VI}}}$ is small for \stellarmass$>10^{9.5}$~\solarmass\space but larger at low masses. For all radial bins, $N_{\mathrm{\ion{O}{VI}}}$ is substantially higher above $10^{9.5}$~\solarmass.\par

\subsection{Radial Profiles of Column Density}
Figure \ref{fig:col_dens_radial} shows the radial dependence of the median and $16^{\mathrm{th}}$ and $84^{\mathrm{th}}$ percentiles for, from top to bottom, $N_{\mathrm{\ion{H}{I}}}$, $N_{\mathrm{\ion{C}{IV}}}$, $N_{\mathrm{\ion{O}{VI}}}$, $N_{\mathrm{\ion{Mg}{II}}}$ and $N_{\mathrm{\ion{Si}{II}}}$ compared with observational data using impact parameter as a proxy for radius. We binned the observational data by stellar mass between $10^{8}$~\solarmass$-10^{11}$~\solarmass\space in equal intervals of $0.5$ dex from left to right. We show the same simulated haloes as in Figure \ref{fig:col_dens_image} which have a stellar mass that is as close to the middle of the stellar mass bin as possible.\par
Our \ion{H}{I} column density shows a steep decrease from ${\sim}0.3$\virialradius. At larger radii, the column density features a shallower decrease out to the virial radius for all haloes up to and including $10^{10.1}$~\solarmass. The scatter of \ion{H}{I} is similar for all haloes up to $10^{10.8}$~\solarmass\space in the inner $0.3$\virialradius. At a stellar mass of $10^{10.8}$~\solarmass, we see significantly wider scatter beyond $0.3$\virialradius\space and shallower decrease than at lower stellar masses out to ${\sim}0.7$\virialradius. The \ion{H}{I} column density in observations is fairly constant between a stellar mass of $10^{8.3}$~\solarmass$-10^{9.3}$~\solarmass, matching our simulated radial profiles at ${\sim}10^{14}\mathrm{cm}^{-2}$. For $10^{9.7}$~\solarmass$\leq$\stellarmass$\leq10^{10.1}$~\solarmass, our simulations underpredict $N_{\mathrm{\ion{H}{I}}}$ outside of 0.5\virialradius\space and match at smaller radii. At $10^{10.8}$~\solarmass, we see better agreement between $0.25$\virialradius$-0.75$\virialradius\space and observations but an overprediction at R$<0.25$\virialradius.\par
\begin{figure*}
\centering
\includegraphics[width=
\textwidth]{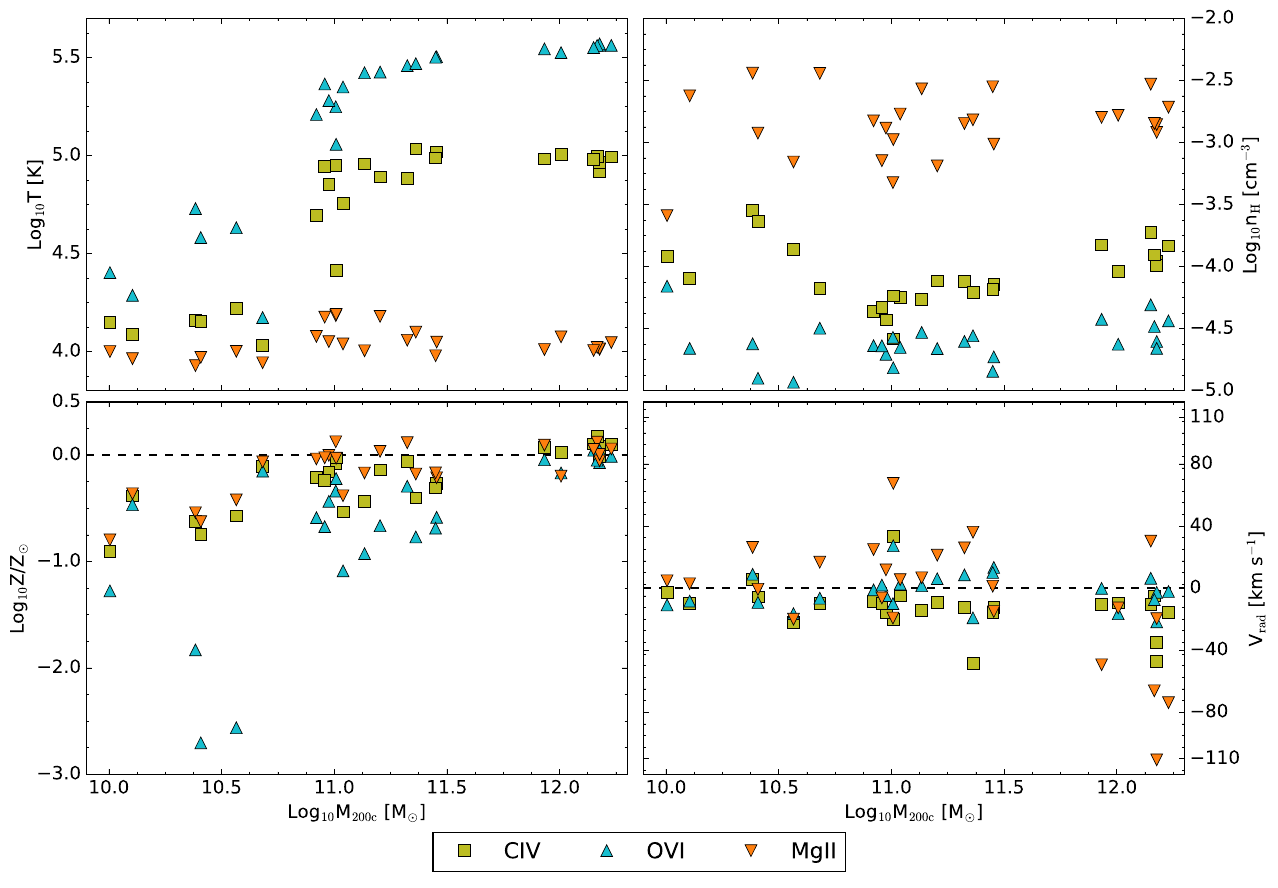}
\vspace{-12pt}
\caption{Temperature (top, left), hydrogen number density (top, right), metallicity (bottom, left) and radial velocity (bottom, right) against \halomass\space as in Fig. \ref{fig:med_properties}. We show the median values weighted by the ion mass of \ion{C}{IV} (yellow squares), \ion{O}{VI} (blue upward triangles) and \ion{Mg}{II} (orange downward triangles) and measured between $0.3\leq{R}/$\virialradius$\leq{1.0}$. \ion{H}{I} and \ion{Si}{II} are similar to \ion{Mg}{II} and are omitted for clarity. We see a sharp increase in temperature of \ion{O}{VI} and \ion{C}{IV} at \halomass$=10^{11}$~\solarmass, indicating a transition from photo-ionization equilibrium to collisional ionization equilibrium. The metallicity is higher in more massive haloes for all ions. The density and radial velocity show no significant dependence on halo mass. There are significant differences between the temperature, the density, the metallicity and the radial velocity of these ions. \ion{Mg}{II} remains at $10^{4}$~K for all haloes, at relatively high densities. \ion{C}{IV} and \ion{O}{VI} trace higher temperatures up to $10^{5}$~K and $10^{5.5}$K, respectively, in more diffuse gas, which explains its ubiquity in the final three rows of Figure \ref{fig:col_dens_image}.}
\label{fig:med_properties_ions}
\end{figure*}
The column densities of \ion{C}{IV} and \ion{O}{VI} steeply decrease with radius for \stellarmass\space$\leq10^{9.3}$~\solarmass. $N_{\mathrm{\ion{C}{IV}}}$ and $N_{\mathrm{\ion{O}{VI}}}$ are higher in the outer CGM in haloes with \halomass\space$\geq 10^{11}$~\solarmass\space or \stellarmass\space$\geq10^{9.7}$~\solarmass. This is likely because of stronger outflows, and because the gas in these haloes has a higher virial temperatures than in smaller haloes. This means there will be more gas at $10^{5}$~K$-10^{5.5}$~K where \ion{C}{IV} and \ion{O}{VI} is commonly found.\par
We see good agreement with observations in our $10^{9.7}$~\solarmass\space stellar mass halo. For \stellarmass$=10^{10.1}$~\solarmass, we find agreement with the limited data available.\par
There are few detections of N$_\mathrm{\ion{O}{VI}}$ at stellar masses below $\leq10^{9.3}$~\solarmass. Our dwarf simulations predict no detectable absorption associated with haloes outside the very centre of these haloes likely a combination of low metallicity because of weak outflows and low temperatures. For \stellarmass$=10^{9.7}$~\solarmass\space and \stellarmass$=10^{10.1}$~\solarmass, our simulations underpredict N$_\mathrm{\ion{O}{VI}}$ in most observations except in the central CGM. At these stellar masses, the \ion{O}{VI} column density decreases with radius more steeply than observations indicate. At $10^{10.8}$~\solarmass, we find the best agreement with detections from observations, but our median and scatter low above most upper limits.\par
For \ion{Mg}{II} and \ion{Si}{II}, the column densities at small radii are high, with a steep decrease between $0.2-0.3$\virialradius\space for \stellarmass\space$\leq10^{10.1}$~\solarmass. We notice a shallower decrease in median and an increase in scatter for both of these ions for \stellarmass\space$=10^{10.8}$~\solarmass, similar to $N_{\mathrm{\ion{H}{I}}}$.\par
Data for stellar masses $\leq10^{9.7}$~\solarmass\space are fairly sparse, leading to a lack of detections to compare with. Above $10^{10.1}$~\solarmass, we find good agreement with detections in the inner $0.25$\virialradius\space for $N_{\mathrm{\ion{Si}{II}}}$. At \stellarmass\space$=10^{10.8}$~\solarmass, we see an overprediction at small radii for both \ion{Mg}{II} and \ion{Si}{II}. We find that our simulations match detections beyond a radius of $0.25$\virialradius\space for \ion{Si}{II}.\par

\subsection{Relation to Physical Properties}
To better understand the properties of gas that give rise to these absorption line features, Figure \ref{fig:med_properties_ions} shows the physical properties as in Figure \ref{fig:med_properties} weighted by the mass of \ion{C}{IV} (yellow squares), \ion{O}{VI} (blue upward triangles) and \ion{Mg}{II} (orange downward arrows). This is to highlight the physical properties traced by the ions we have selected. \ion{H}{I}- and \ion{Si}{II}-weighted values are not shown for clarity, but resemble \ion{Mg}{II} weighted properties. We computed the ion-weighted median values of these properties between $0.3\leq{R}$/\virialradius\space$\leq1.0$. This excludes the central CGM, containing primarily cool, relatively dense, metal-rich and mostly inflowing gas.\par
We see gas traced by \ion{O}{VI} at higher temperatures than \ion{C}{IV}, and gas traced by \ion{Mg}{II} is the coolest, as we would expect for lower ionization states. We see a distinct change in temperature at \halomass\space$=10^{11}$~\solarmass, for \ion{O}{VI} and \ion{C}{IV}. Their temperature increases by ${\sim}0.7$ dex, corresponding to their collisional ionization equilibrium values \citep[see e.g.][]{Strawn2023}. We find that the temperature (density) of \ion{C}{IV} and \ion{O}{VI} lies around ${\sim}10^{5}$~K (${\sim}10^{-4}~\mathrm{cm}^{-3}$) and ${\sim}10^{5.5}$~K (${\sim}10^{-4.8}~\mathrm{cm}^{-3}$), respectively, for \halomass\space$>10^{11}$~\solarmass.\par
\ion{Mg}{II}, as well as \ion{H}{I} and \ion{Si}{II}, are found consistently at low temperatures of ${\sim}10^{4}$K. The densities of the gas probed by these low ionization state ions range from ${\sim}10^{-3.5}-{\sim} 10^{-2.5}~\mathrm{cm^{-3}}$ independent of mass. These ions populate the coolest and densest parts of the outer $70\%$ of the CGM. Additionally, Figure \ref{fig:col_dens_image} shows \ion{Mg}{II} and \ion{Si}{II} in a largely filamentary structure in the most massive halo. By comparing to the most massive halo in in Figure \ref{fig:collated_images}, we can see that filaments found in the halo are at cooler temperatures and higher densities than the rest of the halo, indicating \ion{Mg}{II} and \ion{Si}{II} trace cool and dense filamentary structure.\par
The metallicity of the gas traced by \ion{C}{IV}, \ion{O}{VI} and \ion{Mg}{II} is higher in more massive haloes. \ion{C}{IV} and \ion{Mg}{II} trace higher metallicity gas than \ion{O}{VI} at fixed halo mass. This is because \ion{O}{VI} traces more diffuse gas which is typically more metal poor. Above $\gtrsim10^{12}$~\solarmass, all ions are tracing approximately solar metallicity gas.\par
We find that the ion-weighted radial velocity does not show a strong dependence on halo mass. The median \ion{O}{VI} radial velocity has no preferred direction and is divided equally between slightly inflowing and slightly outflowing gas, while the median \ion{C}{IV} is inflowing for most of our haloes. \ion{Mg}{II} primarily traces outflowing gas at intermediate halo masses ($10^{10.5}$~\solarmass$-10^{11.5}$~\solarmass). \ion{Mg}{II} has previously been found to be dominated by inflows in the inner $60$~kpc of $10^{11.5}$~\solarmass$-10^{12}$~\solarmass\space haloes \citep{DeFelippis2021}. Interestingly, we analyzed the median radial velocity of \ion{Mg}{II} and found inflows dominating above $10^{11.5}$~\solarmass\space in the outer $70\%$ of the CGM.\par

\section{Summary and Discussion}

We have investigated the role the halo mass plays in changing the physical properties of the CGM in the Auriga suite of cosmological simulations. This included an analysis of the temperature, density, metallicity and radial velocity of the CGM for halo masses ranging from $10^{10}$~\solarmass$-10^{12}$~\solarmass. We additionally analysed the column density of \ion{H}{I} and the metal ions \ion{C}{IV}, \ion{O}{VI}, \ion{Mg}{II} and \ion{Si}{II} as a function of stellar mass and radius, and compared to observational data. While this is for one galaxy formation model only, it can give us an idea of what to expect in future observations and can additionally be used to constrain models.\par
We summarise the results of our study below:
\begin{itemize}

\item The CGM of dwarf galaxies (${\sim}10^{10}$~\solarmass) features a narrow radial temperature range. The scatter, defined as the range between the $16^{\mathrm{th}}$ and $84^{\mathrm{th}}$ percentile of our data, in temperature increases from ${\sim}0.5$ dex in $10^{10}$~\solarmass\space haloes to ${\sim}1.5$ dex in $10^{12}$~\solarmass\space haloes. This indicates the multiphase nature of the CGM is dependent on the halo mass.

\item The median and scatter of the volume-weighted density shows no clear dependence on halo mass. The scatter of the mass-weighted density does show some increase with halo mass.

\item The CGM of all of our simulated dwarf galaxies is extremely metal deficient at large radii due to weak outflows up to a halo mass of $10^{11}$~\solarmass. The metallicity in the centre of the CGM of is high but does not reach solar values. The scatter in metallicity is large in within single haloes and from halo-to-halo with metallicities ranging from less than $10^{-5}$~Z$_{\sun}$ up to $10^{-0.5}$~Z$_{\sun}$. Haloes with masses on the order of ${\sim}10^{12}$~\solarmass\space have solar to super-solar metallicity in their centres, and have significantly more metal enriched gas than dwarfs at large radii and exhibit large scatter in metallicity. Most of the gas in Milky Way-mass haloes reaches metallicities between $10-100$\% solar, though there are some lower metallicities above $0.5$\virialradius.

\item We found low outflow velocities in our dwarf galaxies, which means that gas is not efficiently transported out of the inner CGM to larger radii. The gas inflow velocities around dwarf galaxies are much higher. The inflow and outflow velocities increase with halo mass. Above halo masses of ${\sim}10^{11.5}$~\solarmass, the inflow and outflow velocities are roughly equal out to at least the virial radius.

\item The median column densities of our cool regime (\ion{H}{I}, \ion{Mg}{II} and \ion{Si}{II}) are significantly higher than column densities derived from observations in the inner $0.2$\virialradius. $N_{\mathrm{\ion{H}{I}}}$ decreases to between $10^{14}~\mathrm{cm}^{-2}-10^{15}~\mathrm{cm}^{-2}$ at $0.6$\virialradius, in reasonable agreement with observations. N$_\mathrm{\ion{Mg}{II}}$ and N$_\mathrm{\ion{Si}{II}}$ drop off sharply outside of $0.2$\virialradius\space at stellar masses \stellarmass$\leq10^{10}$~\solarmass. There are not enough observational detections to compare to for \ion{Si}{II} below $10^{9.5}$~\solarmass. Above $10^{9.5}$~\solarmass, we find reasonable agreement with detections of \ion{Si}{II} out to $0.4$\virialradius\space and upper limits at larger radii.

\item The median column density of \ion{C}{IV} agrees with observational detections for R$<0.2$\virialradius\space in all our simulated haloes. $N_\mathrm{\ion{C}{IV}}$ decreases at larger radii for all haloes, but shows a very steep decrease in \stellarmass$<10^{9.5}$~\solarmass\space systems, below the few available detections between $0.2$\virialradius$-0.4$\virialradius. The median \ion{O}{VI} column density largely matches observations at high stellar masses (above $10^{9.5}$~\solarmass) but underpredicts them for stellar masses below $10^{9.5}$~\solarmass. The column densities of both \ion{C}{IV} and \ion{O}{VI} in galaxy haloes above a stellar mass of $10^{9.5}$~\solarmass\space decrease by about $1$ dex on average from the centre out to ${\sim}0.6$\virialradius\space which is not seen in observations.

\item We investigated the properties of gas probed by \ion{C}{IV}, \ion{O}{VI} and \ion{Mg}{II} at radii greater than $0.3$\virialradius\space and found that each ion corresponds to its collisional ionization equilibrium temperature of $10^{5}$~K, $10^{5.5}$~K and $10^{4}$~K respectively \citep[e.g. Fig.$2$ from][]{Strawn2023}. A clear transition from photo-ionization equilibrium to collisional ionization equilibrium for \ion{C}{IV} and \ion{O}{VI} is seen at a halo mass of $10^{11}$~\solarmass, featuring a temperature increase from ${\sim}10^{4.1}$~K to ${\sim}10^{5}$~K for \ion{C}{IV} and ${\sim}10^{4.5}$~K to ${\sim}10^{5.5}$~K for \ion{O}{VI}. \ion{Mg}{II} is found in gas that is consistently cool (${\sim}10^{4}$~K) and at densities between  ${\sim}10^{-3.5}~\mathrm{cm}^{-3}-{\sim}10^{-2}~\mathrm{cm}^{-3}$. The ion-weighted metallicity shows that \ion{Mg}{II} and \ion{C}{IV} reside in higher metallicity gas than \ion{O}{VI} for all haloes up to $10^{12}$~\solarmass. The ion-weighted radial velocity shows mostly outflowing \ion{Mg}{II} and inflowing \ion{O}{VI} gas between halo masses of $10^{10.5}$~\solarmass\space and $10^{11.5}$~\solarmass. For halo masses greater than $10^{11.5}$~\solarmass, all metal ions we investigated are preferentially associated with inflowing gas likely tracing large-scale fountain flows.
\end{itemize}
Our resolution tests show similar range scatter across all physical properties irrespective of the resolution of the simulation. In our higher and lower resolution simulations, we see the same trends for the physical properties as a function of halo mass: median temperature (metallicity) increases from $10^{4.5}$~K to ${\sim}10^{6}$~K ($<10^{-5}-10^{0}$) also the scatter increases (decreases) from ${\sim}0.2$ dex to $1.5$ dex (${\sim}4$ dex to $0.3$ dex). The density shows no significant dependence on halo mass nor the median radial velocity. The radial velocity scatter increases from $<10$~km/s to ${\sim}70$km/s with halo mass. Additionally, trends in column density -- increasing with halo mass, decreasing radial profiles and underpredictions compared to observations -- are still present in the level 4 and level 2 simulations. The median and scatter of the column density of all our ions does not deviate much from the level 3 simulations.\par
A similar study by \citet{Hani2019} studied the haloes of 40 L$^{*}$ galaxies in Auriga with stellar mass between $10^{10.3}$~\solarmass$-10^{11.1}$~\solarmass. They found that the gas and metal content of the CGM show a tight correlation with the stellar mass of the host galaxy, similar to our findings. Contrarily, they found the ionization of metals in the CGM is independent of stellar mass, with covering fractions of \ion{H}{I}, \ion{Si}{II}, \ion{C}{IV} and \ion{O}{VI} showing no clear dependence on stellar mass. This seems contradictory to what we find. Haloes of increasing stellar mass feature higher column densities for those ions. However, this may be explained by the smaller stellar mass range ($2.1\leq$\stellarmass$[10^{10}$~\solarmass$]\leq11.7$) and halo mass range ($0.5\leq$\stellarmass$[10^{12}$~\solarmass$]\leq2.0$) of \citet{Hani2019}. Our results show that for a much larger range of stellar masses the column density changes in both median and scatter specifically for high ionization states such as \ion{C}{IV} and \ion{O}{VI}.\par
In a different suite of simulations, called NIHAO \citep{Wang2015}, \citet{Gutcke2017} used a halo mass range similar to our own of $10^{9.7}$~\solarmass$-10^{12.5}$~\solarmass\space and analysed the column density of \ion{H}{I} and \ion{O}{VI}. Using the same self-shielding approximation as our simulations \citep{Rahmati2013}, they found better agreement between their \ion{H}{I} column densities and observations than for \ion{O}{VI}. Additionally, \citet{Gutcke2017} show that the trend of \ion{O}{VI} with luminosity (a proxy for stellar mass) in their simulations agrees with observations but their radial dependence of \ion{O}{VI} underpredicts the observations. In comparison to our work, the column density of \ion{H}{I} follows a similar trend as a function of impact parameter reaching a plateau of ${\sim}10^{14}\mathrm{cm}^{-2}$. We find lower \ion{O}{VI} column densities than \citet{Gutcke2017} out to the virial radius in haloes with \halomass$<10^{11}$~\solarmass. At higher halo masses, \citet{Gutcke2017} find the radial dependence of \ion{O}{VI} has a narrow range of values between $10^{13}\mathrm{cm}^{-2}-10^{14}\mathrm{cm}^{-2}$. We find column densities higher than this, extending to the virial radius.\par
\citet{Hummels2013} conducted similar work investigating the column density of \ion{H}{I}, \ion{Si}{II}, \ion{C}{IV} and \ion{O}{VI} in the CGM of a Milky Way-mass galaxy with different stellar feedback prescriptions. They find that even minimal feedback can still push metals out to large radii ($>50$~kpc) but there is a lack of multiphase gas which underproduces observed ions - this is notably true of \ion{O}{VI} which did not match any of their models. This is different to our results, which show column densities of \ion{O}{VI} in Milky Way mass haloes that better match observations at large radii. They additionally found that increasing thermal feedback from the galaxy to very high values increased the \ion{O}{VI} column density but still underpredicted observations.\par
Similar underpredictions for \ion{O}{VI} and \ion{H}{I} were found in the FIRE-2 suite of MHD simulations from \citet{Ji20} who found that adding cosmic ray feedback in Milky Way mass haloes increased columns in line with observational detections and limits.\par
In order to compute the column density in our simulations, we calculated the metal ion mass fraction in post-processing based on tables generated by \citet{Hummels2017} using \textsc{cloudy} \citep{Ferland2017}. Producing mock spectra for these simulations utilising post-processing tools such as \textsc{trident} \citep{Hummels2017} would allow us to bring our simulation analysis closer to observational data instead of comparing to derived quantities. \citet{Hafen24} used \textsc{trident} to compare synthetic absorption spectra from three different simulation sets. They found reasonable agreement between the density, temperature and metallicity derived from synthetic absorption spectra of uniform clouds and multi-cloud systems, and the source properties used to generate synthetic absorption spectra to within $0.1$ dex. Mock spectra would provide us with a more robust method of comparison between observations and simulations and increase the amount of observational data we can use.\par
Furthermore, previous studies have shown that background or foreground absorbers may be incorrectly attributed to a halo if their line-of-sight velocity falls within the observational velocity range, usually $\pm500$~km~s$^{-1}$, whilst the gas originates from different haloes outside of the virial radius \citep{Ho20,Weng24}. Our column densities were measured along a restricted line-of-sight depth of $2$\virialradius. This could potentially account for discrepancies between our column densities and the observations we compared with, as \citet{Weng24} finds that larger radii from the centre increases the line-of-sight contribution from satellites, other haloes and the IGM.\par
Our results likely depend on specific modelling choices in the Auriga galaxy formation model. Varying certain aspects of our simulations in the future will help to better determine how sensitive our results are to these changes. Potentially relevant subgrid model variations include a kinetic AGN feedback model, cosmic ray feedback, an improved ISM model that incorporates multi-phase gas physics, and a stellar feedback model that is more efficient in dwarf galaxies. Comparisons between CGM predictions and observables are a promising way forward to constrain and distinguish between different galaxy formation models. Additional physical processes such as incorporating cosmic ray feedback in simulations has shown that outflow rates in dwarf galaxies increase and agree more with observations \citep{Dashyan2020,Farcy2022,Defelippis2024}.\par

\section*{Acknowledgements}

FvdV is supported by a Royal Society University Research Fellowship (URF$\backslash$R1$\backslash$191703). RG acknowledges financial support from an STFC Ernest Rutherford Fellowship (ST/W003643/1).
Software used for this work includes \texttt{numpy} \citep{Harris20}, \texttt{matplotlib} \citep{Hunter:2007}. 
The simulations were performed on computing resources provided by the Max Planck Computing and Data Facility in Garching. 

\section*{Data Availability}
 
The Auriga suite of cosmological simulations, including all simulations used in this work, are available at the following website \url{https://wwwmpa.mpa-garching.mpg.de/auriga/dataspecs.html}. The absorption line tables used in this study were part of the \textsc{trident} project and are available at the following website \url{https://trident-project.org/data/ion_table/}.



\bibliographystyle{mnras}
\bibliography{main} 

\begin{thebibliography}{}
\makeatletter
\relax
\def\mn@urlcharsother{\let\do\@makeother \do\$\do\&\do\#\do\^\do\_\do\%\do\~}
\def\mn@doi{\begingroup\mn@urlcharsother \@ifnextchar [ {\mn@doi@} {\mn@doi@[]}}
\def\mn@doi@[#1]#2{\def\@tempa{#1}\ifx\@tempa\@empty \href {http://dx.doi.org/#2} {doi:#2}\else \href {http://dx.doi.org/#2} {#1}\fi \endgroup}
\def\mn@eprint#1#2{\mn@eprint@#1:#2::\@nil}
\def\mn@eprint@arXiv#1{\href {http://arxiv.org/abs/#1} {{\tt arXiv:#1}}}
\def\mn@eprint@dblp#1{\href {http://dblp.uni-trier.de/rec/bibtex/#1.xml} {dblp:#1}}
\def\mn@eprint@#1:#2:#3:#4\@nil{\def\@tempa {#1}\def\@tempb {#2}\def\@tempc {#3}\ifx \@tempc \@empty \let \@tempc \@tempb \let \@tempb \@tempa \fi \ifx \@tempb \@empty \def\@tempb {arXiv}\fi \@ifundefined {mn@eprint@\@tempb}{\@tempb:\@tempc}{\expandafter \expandafter \csname mn@eprint@\@tempb\endcsname \expandafter{\@tempc}}}

\bibitem[\protect\citeauthoryear{{Agertz} et~al.,}{{Agertz} et~al.}{2020}]{Agertz20}
{Agertz} O.,  et~al., 2020, \mn@doi [\mnras] {10.1093/mnras/stz3053}, \href {https://ui.adsabs.harvard.edu/abs/2020MNRAS.491.1656A} {491, 1656}

\bibitem[\protect\citeauthoryear{{Anand}, {Nelson}  \& {Kauffmann}}{{Anand} et~al.}{2021}]{Anand2021}
{Anand} A.,  {Nelson} D.,   {Kauffmann} G.,  2021, \mn@doi [\mnras] {10.1093/mnras/stab871}, \href {https://ui.adsabs.harvard.edu/abs/2021MNRAS.504...65A} {504, 65}

\bibitem[\protect\citeauthoryear{{Appleby}, {Dav{\'e}}, {Sorini}, {Storey-Fisher}  \& {Smith}}{{Appleby} et~al.}{2021}]{Appleby2021}
{Appleby} S.,  {Dav{\'e}} R.,  {Sorini} D.,  {Storey-Fisher} K.,   {Smith} B.,  2021, \mn@doi [\mnras] {10.1093/mnras/stab2310}, \href {https://ui.adsabs.harvard.edu/abs/2021MNRAS.507.2383A} {507, 2383}

\bibitem[\protect\citeauthoryear{{Behroozi}, {Wechsler}  \& {Conroy}}{{Behroozi} et~al.}{2013}]{Behroozi2013}
{Behroozi} P.~S.,  {Wechsler} R.~H.,   {Conroy} C.,  2013, \mn@doi [\apj] {10.1088/0004-637X/770/1/57}, \href {https://ui.adsabs.harvard.edu/abs/2013ApJ...770...57B} {770, 57}

\bibitem[\protect\citeauthoryear{{Bordoloi} et~al.,}{{Bordoloi} et~al.}{2014}]{Bordoloi2014}
{Bordoloi} R.,  et~al., 2014, \mn@doi [\apj] {10.1088/0004-637X/796/2/136}, \href {https://ui.adsabs.harvard.edu/abs/2014ApJ...796..136B} {796, 136}

\bibitem[\protect\citeauthoryear{{Burchett} et~al.,}{{Burchett} et~al.}{2016}]{Burchett2015}
{Burchett} J.~N.,  et~al., 2016, \mn@doi [\apj] {10.3847/0004-637X/832/2/124}, \href {https://ui.adsabs.harvard.edu/abs/2016ApJ...832..124B} {832, 124}

\bibitem[\protect\citeauthoryear{{Dashyan} \& {Dubois}}{{Dashyan} \& {Dubois}}{2020}]{Dashyan2020}
{Dashyan} G.,  {Dubois} Y.,  2020, \mn@doi [\aap] {10.1051/0004-6361/201936339}, \href {https://ui.adsabs.harvard.edu/abs/2020A&A...638A.123D} {638, A123}

\bibitem[\protect\citeauthoryear{{Davis}, {Efstathiou}, {Frenk}  \& {White}}{{Davis} et~al.}{1985}]{Davis1985}
{Davis} M.,  {Efstathiou} G.,  {Frenk} C.~S.,   {White} S.~D.~M.,  1985, \mn@doi [\apj] {10.1086/163168}, \href {https://ui.adsabs.harvard.edu/abs/1985ApJ...292..371D} {292, 371}

\bibitem[\protect\citeauthoryear{{DeFelippis}, {Bouch{\'e}}, {Genel}, {Bryan}, {Nelson}, {Marinacci}  \& {Hernquist}}{{DeFelippis} et~al.}{2021}]{DeFelippis2021}
{DeFelippis} D.,  {Bouch{\'e}} N.~F.,  {Genel} S.,  {Bryan} G.~L.,  {Nelson} D.,  {Marinacci} F.,   {Hernquist} L.,  2021, \mn@doi [\apj] {10.3847/1538-4357/ac2cbf}, \href {https://ui.adsabs.harvard.edu/abs/2021ApJ...923...56D} {923, 56}

\bibitem[\protect\citeauthoryear{{DeFelippis}, {Bournaud}, {Bouch{\'e}}, {Tollet}, {Farcy}, {Rey}, {Rosdahl}  \& {Blaizot}}{{DeFelippis} et~al.}{2024}]{Defelippis2024}
{DeFelippis} D.,  {Bournaud} F.,  {Bouch{\'e}} N.,  {Tollet} E.,  {Farcy} M.,  {Rey} M.,  {Rosdahl} J.,   {Blaizot} J.,  2024, \mn@doi [\mnras] {10.1093/mnras/stae837}, \href {https://ui.adsabs.harvard.edu/abs/2024MNRAS.530...52D} {530, 52}

\bibitem[\protect\citeauthoryear{{Farcy}, {Rosdahl}, {Dubois}, {Blaizot}  \& {Martin-Alvarez}}{{Farcy} et~al.}{2022}]{Farcy2022}
{Farcy} M.,  {Rosdahl} J.,  {Dubois} Y.,  {Blaizot} J.,   {Martin-Alvarez} S.,  2022, \mn@doi [\mnras] {10.1093/mnras/stac1196}, \href {https://ui.adsabs.harvard.edu/abs/2022MNRAS.513.5000F} {513, 5000}

\bibitem[\protect\citeauthoryear{{Faucher-Gigu{\`e}re}, {Lidz}, {Zaldarriaga}  \& {Hernquist}}{{Faucher-Gigu{\`e}re} et~al.}{2009}]{Giguere2009}
{Faucher-Gigu{\`e}re} C.-A.,  {Lidz} A.,  {Zaldarriaga} M.,   {Hernquist} L.,  2009, \mn@doi [\apj] {10.1088/0004-637X/703/2/1416}, \href {https://ui.adsabs.harvard.edu/abs/2009ApJ...703.1416F} {703, 1416}

\bibitem[\protect\citeauthoryear{{Ferland} et~al.,}{{Ferland} et~al.}{2017}]{Ferland2017}
{Ferland} G.~J.,  et~al., 2017, \mn@doi [\rmxaa] {10.48550/arXiv.1705.10877}, \href {https://ui.adsabs.harvard.edu/abs/2017RMxAA..53..385F} {53, 385}

\bibitem[\protect\citeauthoryear{{Fox}, {Wakker}, {Savage}, {Tripp}, {Sembach}  \& {Bland-Hawthorn}}{{Fox} et~al.}{2005}]{Fox2005}
{Fox} A.~J.,  {Wakker} B.~P.,  {Savage} B.~D.,  {Tripp} T.~M.,  {Sembach} K.~R.,   {Bland-Hawthorn} J.,  2005, \mn@doi [\apj] {10.1086/431915}, \href {https://ui.adsabs.harvard.edu/abs/2005ApJ...630..332F} {630, 332}

\bibitem[\protect\citeauthoryear{{Frenk}, {White}, {Davis}  \& {Efstathiou}}{{Frenk} et~al.}{1988}]{Frenk1988}
{Frenk} C.~S.,  {White} S. D.~M.,  {Davis} M.,   {Efstathiou} G.,  1988, \mn@doi [\apj] {10.1086/166213}, \href {https://ui.adsabs.harvard.edu/abs/1988ApJ...327..507F} {327, 507}

\bibitem[\protect\citeauthoryear{{Girelli}, {Pozzetti}, {Bolzonella}, {Giocoli}, {Marulli}  \& {Baldi}}{{Girelli} et~al.}{2020}]{Girelli2020}
{Girelli} G.,  {Pozzetti} L.,  {Bolzonella} M.,  {Giocoli} C.,  {Marulli} F.,   {Baldi} M.,  2020, \mn@doi [\aap] {10.1051/0004-6361/201936329}, \href {https://ui.adsabs.harvard.edu/abs/2020A&A...634A.135G} {634, A135}

\bibitem[\protect\citeauthoryear{{Gnat} \& {Sternberg}}{{Gnat} \& {Sternberg}}{2007}]{Gnat2007}
{Gnat} O.,  {Sternberg} A.,  2007, \mn@doi [\apjs] {10.1086/509786}, \href {https://ui.adsabs.harvard.edu/abs/2007ApJS..168..213G} {168, 213}

\bibitem[\protect\citeauthoryear{{Grand} et~al.,}{{Grand} et~al.}{2017}]{Grand2017}
{Grand} R. J.~J.,  et~al., 2017, \mn@doi [\mnras] {10.1093/mnras/stx071}, \href {https://ui.adsabs.harvard.edu/abs/2017MNRAS.467..179G} {467, 179}

\bibitem[\protect\citeauthoryear{{Grand}, {Fragkoudi}, {G{\'o}mez}, {Jenkins}, {Marinacci}, {Pakmor}  \& {Springel}}{{Grand} et~al.}{2024}]{Grand2024}
{Grand} R. J.~J.,  {Fragkoudi} F.,  {G{\'o}mez} F.~A.,  {Jenkins} A.,  {Marinacci} F.,  {Pakmor} R.,   {Springel} V.,  2024, \mn@doi [arXiv e-prints] {10.48550/arXiv.2401.08750}, \href {https://ui.adsabs.harvard.edu/abs/2024arXiv240108750G} {p. arXiv:2401.08750}

\bibitem[\protect\citeauthoryear{{Gutcke}, {Stinson}, {Macci{\`o}}, {Wang}  \& {Dutton}}{{Gutcke} et~al.}{2017}]{Gutcke2017}
{Gutcke} T.~A.,  {Stinson} G.~S.,  {Macci{\`o}} A.~V.,  {Wang} L.,   {Dutton} A.~A.,  2017, \mn@doi [\mnras] {10.1093/mnras/stw2539}, \href {https://ui.adsabs.harvard.edu/abs/2017MNRAS.464.2796G} {464, 2796}

\bibitem[\protect\citeauthoryear{{Gutcke}, {Pakmor}, {Naab}  \& {Springel}}{{Gutcke} et~al.}{2021}]{Gutcke2021}
{Gutcke} T.~A.,  {Pakmor} R.,  {Naab} T.,   {Springel} V.,  2021, \mn@doi [\mnras] {10.1093/mnras/staa3875}, \href {https://ui.adsabs.harvard.edu/abs/2021MNRAS.501.5597G} {501, 5597}

\bibitem[\protect\citeauthoryear{{Hafen} et~al.,}{{Hafen} et~al.}{2020}]{Hafen2020}
{Hafen} Z.,  et~al., 2020, \mn@doi [\mnras] {10.1093/mnras/staa902}, \href {https://ui.adsabs.harvard.edu/abs/2020MNRAS.494.3581H} {494, 3581}

\bibitem[\protect\citeauthoryear{{Hafen} et~al.,}{{Hafen} et~al.}{2024}]{Hafen24}
{Hafen} Z.,  et~al., 2024, \mn@doi [\mnras] {10.1093/mnras/stad3889}, \href {https://ui.adsabs.harvard.edu/abs/2024MNRAS.528...39H} {528, 39}

\bibitem[\protect\citeauthoryear{{Hani}, {Ellison}, {Sparre}, {Grand}, {Pakmor}, {Gomez}  \& {Springel}}{{Hani} et~al.}{2019}]{Hani2019}
{Hani} M.~H.,  {Ellison} S.~L.,  {Sparre} M.,  {Grand} R. J.~J.,  {Pakmor} R.,  {Gomez} F.~A.,   {Springel} V.,  2019, \mn@doi [\mnras] {10.1093/mnras/stz1708}, \href {https://ui.adsabs.harvard.edu/abs/2019MNRAS.488..135H} {488, 135}

\bibitem[\protect\citeauthoryear{Harris, Millman  \& van~der Walt}{Harris et~al.}{2020}]{Harris20}
Harris C.,  Millman K.,   van~der Walt S. e.~a.,  2020, \mn@doi [Nature] {doi.org/10.1038/s41586-020-2649-2}, 582, 357

\bibitem[\protect\citeauthoryear{{Ho}, {Martin}  \& {Schaye}}{{Ho} et~al.}{2020}]{Ho20}
{Ho} S.~H.,  {Martin} C.~L.,   {Schaye} J.,  2020, \mn@doi [\apj] {10.3847/1538-4357/abbe88}, \href {https://ui.adsabs.harvard.edu/abs/2020ApJ...904...76H} {904, 76}

\bibitem[\protect\citeauthoryear{{Hummels}, {Bryan}, {Smith}  \& {Turk}}{{Hummels} et~al.}{2013}]{Hummels2013}
{Hummels} C.~B.,  {Bryan} G.~L.,  {Smith} B.~D.,   {Turk} M.~J.,  2013, \mn@doi [\mnras] {10.1093/mnras/sts702}, \href {https://ui.adsabs.harvard.edu/abs/2013MNRAS.430.1548H} {430, 1548}

\bibitem[\protect\citeauthoryear{{Hummels}, {Smith}  \& {Silvia}}{{Hummels} et~al.}{2017}]{Hummels2017}
{Hummels} C.~B.,  {Smith} B.~D.,   {Silvia} D.~W.,  2017, \mn@doi [\apj] {10.3847/1538-4357/aa7e2d}, \href {https://ui.adsabs.harvard.edu/abs/2017ApJ...847...59H} {847, 59}

\bibitem[\protect\citeauthoryear{Hunter}{Hunter}{2007}]{Hunter:2007}
Hunter J.~D.,  2007, \mn@doi [Computing in Science \& Engineering] {10.1109/MCSE.2007.55}, 9, 90

\bibitem[\protect\citeauthoryear{{Ji} et~al.,}{{Ji} et~al.}{2020}]{Ji20}
{Ji} S.,  et~al., 2020, \mn@doi [\mnras] {10.1093/mnras/staa1849}, \href {https://ui.adsabs.harvard.edu/abs/2020MNRAS.496.4221J} {496, 4221}

\bibitem[\protect\citeauthoryear{{Johnson}, {Chen}  \& {Mulchaey}}{{Johnson} et~al.}{2015}]{Johnson2015}
{Johnson} S.~D.,  {Chen} H.-W.,   {Mulchaey} J.~S.,  2015, \mn@doi [\mnras] {10.1093/mnras/stv553}, \href {https://ui.adsabs.harvard.edu/abs/2015MNRAS.449.3263J} {449, 3263}

\bibitem[\protect\citeauthoryear{{Johnson}, {Chen}, {Mulchaey}, {Schaye}  \& {Straka}}{{Johnson} et~al.}{2017}]{Johnson2017}
{Johnson} S.~D.,  {Chen} H.-W.,  {Mulchaey} J.~S.,  {Schaye} J.,   {Straka} L.~A.,  2017, \mn@doi [\apjl] {10.3847/2041-8213/aa9370}, \href {https://ui.adsabs.harvard.edu/abs/2017ApJ...850L..10J} {850, L10}

\bibitem[\protect\citeauthoryear{{Kannan}, {Vogelsberger}, {Marinacci}, {McKinnon}, {Pakmor}  \& {Springel}}{{Kannan} et~al.}{2019}]{Kannan2019}
{Kannan} R.,  {Vogelsberger} M.,  {Marinacci} F.,  {McKinnon} R.,  {Pakmor} R.,   {Springel} V.,  2019, \mn@doi [\mnras] {10.1093/mnras/stz287}, \href {https://ui.adsabs.harvard.edu/abs/2019MNRAS.485..117K} {485, 117}

\bibitem[\protect\citeauthoryear{{Katz}}{{Katz}}{2022}]{Katz2022}
{Katz} H.,  2022, \mn@doi [\mnras] {10.1093/mnras/stac423}, \href {https://ui.adsabs.harvard.edu/abs/2022MNRAS.512..348K} {512, 348}

\bibitem[\protect\citeauthoryear{{Kere{\v{s}}}, {Katz}, {Weinberg}  \& {Dav{\'e}}}{{Kere{\v{s}}} et~al.}{2005}]{Keres2005}
{Kere{\v{s}}} D.,  {Katz} N.,  {Weinberg} D.~H.,   {Dav{\'e}} R.,  2005, \mn@doi [\mnras] {10.1111/j.1365-2966.2005.09451.x}, \href {https://ui.adsabs.harvard.edu/abs/2005MNRAS.363....2K} {363, 2}

\bibitem[\protect\citeauthoryear{{Lehner} et~al.,}{{Lehner} et~al.}{2013}]{Lehner2013}
{Lehner} N.,  et~al., 2013, \mn@doi [\apj] {10.1088/0004-637X/770/2/138}, \href {https://ui.adsabs.harvard.edu/abs/2013ApJ...770..138L} {770, 138}

\bibitem[\protect\citeauthoryear{{Machado}, {Tissera}, {Lima Neto}  \& {Sodr{\'e}}}{{Machado} et~al.}{2018}]{Machado2018}
{Machado} R.~E.~G.,  {Tissera} P.~B.,  {Lima Neto} G.~B.,   {Sodr{\'e}} L.,  2018, \mn@doi [\aap] {10.1051/0004-6361/201628886}, \href {https://ui.adsabs.harvard.edu/abs/2018A&A...609A..66M} {609, A66}

\bibitem[\protect\citeauthoryear{{Mathur}, {Gupta}, {Das}, {Krongold}  \& {Nicastro}}{{Mathur} et~al.}{2021}]{Mathur2021}
{Mathur} S.,  {Gupta} A.,  {Das} S.,  {Krongold} Y.,   {Nicastro} F.,  2021, \mn@doi [\apj] {10.3847/1538-4357/abd03f}, \href {https://ui.adsabs.harvard.edu/abs/2021ApJ...908...69M} {908, 69}

\bibitem[\protect\citeauthoryear{{Mo}, {van den Bosch}  \& {White}}{{Mo} et~al.}{2010}]{Mo2010}
{Mo} H.,  {van den Bosch} F.~C.,   {White} S.,  2010, {Galaxy Formation and Evolution}.
Cambridge University Press

\bibitem[\protect\citeauthoryear{{Moster}, {Somerville}, {Maulbetsch}, {van den Bosch}, {Macci{\`o}}, {Naab}  \& {Oser}}{{Moster} et~al.}{2010}]{Moster2010}
{Moster} B.~P.,  {Somerville} R.~S.,  {Maulbetsch} C.,  {van den Bosch} F.~C.,  {Macci{\`o}} A.~V.,  {Naab} T.,   {Oser} L.,  2010, \mn@doi [\apj] {10.1088/0004-637X/710/2/903}, \href {https://ui.adsabs.harvard.edu/abs/2010ApJ...710..903M} {710, 903}

\bibitem[\protect\citeauthoryear{{Muratov}, {Kere{\v{s}}}, {Faucher-Gigu{\`e}re}, {Hopkins}, {Quataert}  \& {Murray}}{{Muratov} et~al.}{2015}]{Muratov2015}
{Muratov} A.~L.,  {Kere{\v{s}}} D.,  {Faucher-Gigu{\`e}re} C.-A.,  {Hopkins} P.~F.,  {Quataert} E.,   {Murray} N.,  2015, \mn@doi [\mnras] {10.1093/mnras/stv2126}, \href {https://ui.adsabs.harvard.edu/abs/2015MNRAS.454.2691M} {454, 2691}

\bibitem[\protect\citeauthoryear{{Oppenheimer}, {Dav{\'e}}, {Kere{\v{s}}}, {Fardal}, {Katz}, {Kollmeier}  \& {Weinberg}}{{Oppenheimer} et~al.}{2010}]{Oppenheimer2010}
{Oppenheimer} B.~D.,  {Dav{\'e}} R.,  {Kere{\v{s}}} D.,  {Fardal} M.,  {Katz} N.,  {Kollmeier} J.~A.,   {Weinberg} D.~H.,  2010, \mn@doi [\mnras] {10.1111/j.1365-2966.2010.16872.x}, \href {https://ui.adsabs.harvard.edu/abs/2010MNRAS.406.2325O} {406, 2325}

\bibitem[\protect\citeauthoryear{{Pakmor}, {Springel}, {Bauer}, {Mocz}, {Munoz}, {Ohlmann}, {Schaal}  \& {Zhu}}{{Pakmor} et~al.}{2016}]{Pakmor2016a}
{Pakmor} R.,  {Springel} V.,  {Bauer} A.,  {Mocz} P.,  {Munoz} D.~J.,  {Ohlmann} S.~T.,  {Schaal} K.,   {Zhu} C.,  2016, \mn@doi [\mnras] {10.1093/mnras/stv2380}, \href {https://ui.adsabs.harvard.edu/abs/2016MNRAS.455.1134P} {455, 1134}

\bibitem[\protect\citeauthoryear{{Peeples}, {Werk}, {Tumlinson}, {Oppenheimer}, {Prochaska}, {Katz}  \& {Weinberg}}{{Peeples} et~al.}{2014}]{Peeples2014}
{Peeples} M.~S.,  {Werk} J.~K.,  {Tumlinson} J.,  {Oppenheimer} B.~D.,  {Prochaska} J.~X.,  {Katz} N.,   {Weinberg} D.~H.,  2014, \mn@doi [\apj] {10.1088/0004-637X/786/1/54}, \href {https://ui.adsabs.harvard.edu/abs/2014ApJ...786...54P} {786, 54}

\bibitem[\protect\citeauthoryear{{Peeples} et~al.,}{{Peeples} et~al.}{2019}]{Peeples2019}
{Peeples} M.,  et~al., 2019, \mn@doi [\baas] {10.48550/arXiv.1903.05644}, \href {https://ui.adsabs.harvard.edu/abs/2019BAAS...51c.368P} {51, 368}

\bibitem[\protect\citeauthoryear{{Planck Collaboration} et~al.,}{{Planck Collaboration} et~al.}{2014}]{Planck2014}
{Planck Collaboration} et~al., 2014, \mn@doi [\aap] {10.1051/0004-6361/201321591}, \href {https://ui.adsabs.harvard.edu/abs/2014A&A...571A..16P} {571, A16}

\bibitem[\protect\citeauthoryear{{Portinari}, {Chiosi}  \& {Bressan}}{{Portinari} et~al.}{1998}]{Portinari1998}
{Portinari} L.,  {Chiosi} C.,   {Bressan} A.,  1998, \mn@doi [\aap] {10.48550/arXiv.astro-ph/9711337}, \href {https://ui.adsabs.harvard.edu/abs/1998A&A...334..505P} {334, 505}

\bibitem[\protect\citeauthoryear{{Rahmati}, {Pawlik}, {Rai{\v{c}}evi{\'c}}  \& {Schaye}}{{Rahmati} et~al.}{2013}]{Rahmati2013}
{Rahmati} A.,  {Pawlik} A.~H.,  {Rai{\v{c}}evi{\'c}} M.,   {Schaye} J.,  2013, \mn@doi [\mnras] {10.1093/mnras/stt066}, \href {https://ui.adsabs.harvard.edu/abs/2013MNRAS.430.2427R} {430, 2427}

\bibitem[\protect\citeauthoryear{{Ranjan} et~al.,}{{Ranjan} et~al.}{2022}]{Ranjan_2022}
{Ranjan} A.,  et~al., 2022, \mn@doi [\aap] {10.1051/0004-6361/202140604}, \href {https://ui.adsabs.harvard.edu/abs/2022A&A...661A.134R} {661, A134}

\bibitem[\protect\citeauthoryear{{Sales}, {Wetzel}  \& {Fattahi}}{{Sales} et~al.}{2022}]{Sales22}
{Sales} L.~V.,  {Wetzel} A.,   {Fattahi} A.,  2022, \mn@doi [Nature Astronomy] {10.1038/s41550-022-01689-w}, \href {https://ui.adsabs.harvard.edu/abs/2022NatAs...6..897S} {6, 897}

\bibitem[\protect\citeauthoryear{{Sanchez}, {Werk}, {Tremmel}, {Pontzen}, {Christensen}, {Quinn}  \& {Cruz}}{{Sanchez} et~al.}{2019}]{Sanchez2019}
{Sanchez} N.~N.,  {Werk} J.~K.,  {Tremmel} M.,  {Pontzen} A.,  {Christensen} C.,  {Quinn} T.,   {Cruz} A.,  2019, \mn@doi [\apj] {10.3847/1538-4357/ab3045}, \href {https://ui.adsabs.harvard.edu/abs/2019ApJ...882....8S} {882, 8}

\bibitem[\protect\citeauthoryear{{Schaye} et~al.,}{{Schaye} et~al.}{2015}]{Schaye2015}
{Schaye} J.,  et~al., 2015, \mn@doi [\mnras] {10.1093/mnras/stu2058}, \href {https://ui.adsabs.harvard.edu/abs/2015MNRAS.446..521S} {446, 521}

\bibitem[\protect\citeauthoryear{{Springel}}{{Springel}}{2010}]{Springel2010}
{Springel} V.,  2010, \mn@doi [\mnras] {10.1111/j.1365-2966.2009.15715.x}, \href {https://ui.adsabs.harvard.edu/abs/2010MNRAS.401..791S} {401, 791}

\bibitem[\protect\citeauthoryear{{Springel} \& {Hernquist}}{{Springel} \& {Hernquist}}{2003}]{Springel2003}
{Springel} V.,  {Hernquist} L.,  2003, \mn@doi [\mnras] {10.1046/j.1365-8711.2003.06206.x}, \href {https://ui.adsabs.harvard.edu/abs/2003MNRAS.339..289S} {339, 289}

\bibitem[\protect\citeauthoryear{{Strawn}, {Roca-F{\`a}brega}  \& {Primack}}{{Strawn} et~al.}{2023}]{Strawn2023}
{Strawn} C.,  {Roca-F{\`a}brega} S.,   {Primack} J.,  2023, \mn@doi [\mnras] {10.1093/mnras/stac3567}, \href {https://ui.adsabs.harvard.edu/abs/2023MNRAS.519....1S} {519, 1}

\bibitem[\protect\citeauthoryear{{Tchernyshyov} et~al.,}{{Tchernyshyov} et~al.}{2022}]{Tchernyshyov2022}
{Tchernyshyov} K.,  et~al., 2022, \mn@doi [\apj] {10.3847/1538-4357/ac450c}, \href {https://ui.adsabs.harvard.edu/abs/2022ApJ...927..147T} {927, 147}

\bibitem[\protect\citeauthoryear{{Tumlinson} et~al.,}{{Tumlinson} et~al.}{2011}]{Tumlinson2011}
{Tumlinson} J.,  et~al., 2011, \mn@doi [Science] {10.1126/science.1209840}, \href {https://ui.adsabs.harvard.edu/abs/2011Sci...334..948T} {334, 948}

\bibitem[\protect\citeauthoryear{{Tumlinson} et~al.,}{{Tumlinson} et~al.}{2013}]{Tumlinson2013}
{Tumlinson} J.,  et~al., 2013, \mn@doi [\apj] {10.1088/0004-637X/777/1/59}, \href {https://ui.adsabs.harvard.edu/abs/2013ApJ...777...59T} {777, 59}

\bibitem[\protect\citeauthoryear{{Tumlinson}, {Peeples}  \& {Werk}}{{Tumlinson} et~al.}{2017}]{Tumlinson2017}
{Tumlinson} J.,  {Peeples} M.~S.,   {Werk} J.~K.,  2017, \mn@doi [\araa] {10.1146/annurev-astro-091916-055240}, \href {https://ui.adsabs.harvard.edu/abs/2017ARA&A..55..389T} {55, 389}

\bibitem[\protect\citeauthoryear{{Vogelsberger}, {Genel}, {Sijacki}, {Torrey}, {Springel}  \& {Hernquist}}{{Vogelsberger} et~al.}{2013}]{vogel2013}
{Vogelsberger} M.,  {Genel} S.,  {Sijacki} D.,  {Torrey} P.,  {Springel} V.,   {Hernquist} L.,  2013, \mn@doi [\mnras] {10.1093/mnras/stt1789}, \href {https://ui.adsabs.harvard.edu/abs/2013MNRAS.436.3031V} {436, 3031}

\bibitem[\protect\citeauthoryear{{Wang} \& {Abel}}{{Wang} \& {Abel}}{2008}]{Wang2008}
{Wang} P.,  {Abel} T.,  2008, \mn@doi [\apj] {10.1086/523623}, \href {https://ui.adsabs.harvard.edu/abs/2008ApJ...672..752W} {672, 752}

\bibitem[\protect\citeauthoryear{{Wang}, {Dutton}, {Stinson}, {Macci{\`o}}, {Penzo}, {Kang}, {Keller}  \& {Wadsley}}{{Wang} et~al.}{2015}]{Wang2015}
{Wang} L.,  {Dutton} A.~A.,  {Stinson} G.~S.,  {Macci{\`o}} A.~V.,  {Penzo} C.,  {Kang} X.,  {Keller} B.~W.,   {Wadsley} J.,  2015, \mn@doi [\mnras] {10.1093/mnras/stv1937}, \href {https://ui.adsabs.harvard.edu/abs/2015MNRAS.454...83W} {454, 83}

\bibitem[\protect\citeauthoryear{{Weng}, {P{\'e}roux}, {Ramesh}, {Nelson}, {Sadler}, {Zwaan}, {Bollo}  \& {Casavecchia}}{{Weng} et~al.}{2024}]{Weng24}
{Weng} S.,  {P{\'e}roux} C.,  {Ramesh} R.,  {Nelson} D.,  {Sadler} E.~M.,  {Zwaan} M.,  {Bollo} V.,   {Casavecchia} B.,  2024, \mn@doi [\mnras] {10.1093/mnras/stad3426}, \href {https://ui.adsabs.harvard.edu/abs/2024MNRAS.527.3494W} {527, 3494}

\bibitem[\protect\citeauthoryear{{Werk}, {Prochaska}, {Thom}, {Tumlinson}, {Tripp}, {O'Meara}  \& {Peeples}}{{Werk} et~al.}{2013}]{Werk2013}
{Werk} J.~K.,  {Prochaska} J.~X.,  {Thom} C.,  {Tumlinson} J.,  {Tripp} T.~M.,  {O'Meara} J.~M.,   {Peeples} M.~S.,  2013, \mn@doi [\apjs] {10.1088/0067-0049/204/2/17}, \href {https://ui.adsabs.harvard.edu/abs/2013ApJS..204...17W} {204, 17}

\bibitem[\protect\citeauthoryear{{Werk} et~al.,}{{Werk} et~al.}{2016}]{Werk2016}
{Werk} J.~K.,  et~al., 2016, \mn@doi [\apj] {10.3847/1538-4357/833/1/54}, \href {https://ui.adsabs.harvard.edu/abs/2016ApJ...833...54W} {833, 54}

\bibitem[\protect\citeauthoryear{{Wiersma}, {Schaye}  \& {Smith}}{{Wiersma} et~al.}{2009}]{Wiersma2009}
{Wiersma} R. P.~C.,  {Schaye} J.,   {Smith} B.~D.,  2009, \mn@doi [\mnras] {10.1111/j.1365-2966.2008.14191.x}, \href {https://ui.adsabs.harvard.edu/abs/2009MNRAS.393...99W} {393, 99}

\bibitem[\protect\citeauthoryear{{Wright}, {Lagos}, {Power}  \& {Correa}}{{Wright} et~al.}{2021}]{Wright2021}
{Wright} R.~J.,  {Lagos} C. d.~P.,  {Power} C.,   {Correa} C.~A.,  2021, \mn@doi [\mnras] {10.1093/mnras/stab1057}, \href {https://ui.adsabs.harvard.edu/abs/2021MNRAS.504.5702W} {504, 5702}

\bibitem[\protect\citeauthoryear{{Zheng}, {Emerick}, {Putman}, {Werk}, {Kirby}  \& {Peek}}{{Zheng} et~al.}{2020}]{Zheng2020}
{Zheng} Y.,  {Emerick} A.,  {Putman} M.~E.,  {Werk} J.~K.,  {Kirby} E.~N.,   {Peek} J.,  2020, \mn@doi [\apj] {10.3847/1538-4357/abc875}, \href {https://ui.adsabs.harvard.edu/abs/2020ApJ...905..133Z} {905, 133}

\bibitem[\protect\citeauthoryear{{Zheng} et~al.,}{{Zheng} et~al.}{2024}]{Zheng2023}
{Zheng} Y.,  et~al., 2024, \mn@doi [\apj] {10.3847/1538-4357/acfe6b}, \href {https://ui.adsabs.harvard.edu/abs/2024ApJ...960...55Z} {960, 55}

\bibitem[\protect\citeauthoryear{{van de Voort} \& {Schaye.}}{{van de Voort} \& {Schaye.}}{2012}]{vandevoort2012}
{van de Voort} F.,  {Schaye.} J.,  2012, \mn@doi [\mnras] {10.1111/j.1365-2966.2012.20949.x}, \href {https://ui.adsabs.harvard.edu/abs/2012MNRAS.423.2991V} {423, 2991}

\makeatother
\end{thebibliography}


\newpage
\appendix

\section{Halo and Galaxy Properties}

In Table \ref{table:1}, we outline global properties from the most massive halo in each of our simulations, selected to be relatively isolated. These data outline the halo name, virial mass, virial temperature, virial radius, star formation rate and stellar mass of the most-massive halo within the zoom-in region of each simulation.

\begin{table*}
    \begin{tabular}{|c|c|c|c|c|c|}
    \hline\hline
        Halo ID$^{a}$ & Log$_{10}$M$_{200c}^{b}$ [M$_{\sun}$] &  T$_{vir}^{c}$ [K] &
 R$_{200c}^{d}$ (kpc) & SFR$^{e}$ [$\mathrm{{M}_{\sun}}/\mathrm{yr}$] &
Log$_{10}$M$_{\star}^{f}$ [M$_{\sun}$] \\ \hline\multicolumn{6}{c}{\halomass$=10^{10}$~\solarmass$-10^{11}$~\solarmass\space Haloes}\\\hline
        halo\_0 & 10.00 & 4.35 & 45.58 & 0.00 & 7.46 \\
        halo\_2 & 10.68 & 4.84 & 76.67 & 0.46 & 9.36 \\
        halo\_6 & 10.38 & 4.61 & 61.00 & 0.10 & 8.48 \\
        halo\_8 & 10.10 & 4.46 & 49.22 & 0.05 & 8.48 \\
        halo\_9 & 10.56 & 4.78 & 70.10 & 0.16 & 8.85 \\
        halo\_11 & 10.40 & 4.65 & 62.11 & 0.08 & 8.34 \\ \hline\multicolumn{6}{c}{\halomass$={\sim}10^{11}$~\solarmass$-10^{12}$~\solarmass\space Haloes}\\\hline
        halo\_0 & 11.00 & 4.89 & 98.36 & 0.79 & 9.77 \\
        halo\_1 & 11.32 & 5.14 & 125.47 & 2.06 & 10.08 \\
        halo\_2 & 11.13 & 5.11 & 108.48 & 0.78 & 9.74 \\
        halo\_3 & 11.36 & 5.21 & 129.18 & 1.15 & 10.08 \\
        halo\_4 & 11.45 & 5.33 & 138.67 & 1.43 & 10.08 \\
        halo\_5 & 11.45 & 5.35 & 138.17 & 0.60 & 10.14 \\
        halo\_6 & 10.97 & 4.94 & 96.06 & 0.52 & 9.72 \\
        halo\_7 & 11.20 & 5.07 & 114.38 & 1.41 & 9.92 \\
        halo\_8 & 11.03 & 5.06 & 100.82 & 0.46 & 9.53 \\
        halo\_9 & 11.00 & 4.89 & 98.47 & 0.62 & 9.63 \\
        halo\_10 & 10.92 & 4.88 & 92.07 & 0.47 & 9.55 \\
        halo\_11 & 10.95 & 4.98 & 94.72 & 0.89 & 9.67 \\ \hline\multicolumn{6}{c}{\halomass$\gtrsim10^{12}$~\solarmass\space Haloes}\\\hline
        halo\_L8 & 11.93 & 5.57 & 200.20 & 5.71 & 10.82\\
        halo\_6 & 12.00 & 5.58 & 211.83 & 2.71 & 10.81 \\
        halo\_16 & 12.18 & 5.81 & 241.53 & 4.20 & 10.96 \\
        halo\_21 & 12.15 & 5.71 & 236.69 & 10.27 & 10.96 \\
        halo\_23 & 12.17 & 5.71 & 241.50 & 4.60 & 10.97 \\
        halo\_24 & 12.17 & 5.76 & 239.57 & 7.67 & 10.95 \\
        halo\_27 & 12.23 & 5.70 & 251.40 & 4.20 & 11.01 \\\hline\hline
    \end{tabular}
    \caption[size=8pt]{Properties of our most massive haloes in each of our simulations. a) simulation reference name b) halo mass c) virial temperature d) virial radius of halo e) stellar mass f) star formation rate.}
    \label{table:1}
\end{table*}

\bsp	
\label{lastpage}
\end{document}